%% file: main.tex
\documentclass[pra,twocolumn,superscriptaddress,amsmath,amssymb,longbibliography,floatfix]{revtex4-2}

\usepackage{graphicx}
\usepackage[colorlinks=true,linkcolor=blue,citecolor=blue,urlcolor=blue]{hyperref}
\usepackage{cleveref}
\makeatletter
\expandafter\gdef\csname cref@@format\endcsname#1#2#3{#2Sec.~#1#3}
\expandafter\gdef\csname Cref@@format\endcsname#1#2#3{#2Section~#1#3}
\expandafter\gdef\csname crefrange@@format\endcsname#1#2#3#4#5#6{%
  Secs.~#1#2 to~#4#5#6}
\expandafter\gdef\csname Crefrange@@format\endcsname#1#2#3#4#5#6{%
  Sections~#1#2 to~#4#5#6}
\makeatother
\crefformat{section}{#2Sec.~#1#3}
\Crefformat{section}{#2Section~#1#3}
\crefname{section}{Sec.}{Secs.}
\Crefname{section}{Section}{Sections}
\usepackage{booktabs}
\usepackage{siunitx}
\usepackage{multirow}
\usepackage{tabularx}
\usepackage{xcolor}
\usepackage{enumitem}
\usepackage{placeins}   

\newcommand{\vect}[1]{\mathbf{#1}}
\newcommand{\mat}[1]{\mathbf{#1}}

\newcommand{\prism}{\textsc{Prism}}

\begin{document}

\title{PRISM: Photonic Similarity Engine for KV Cache Block Selection\\in Long-Context LLM Inference}

\author{Hyoseok Park}
\affiliation{Department of Physics, Chungnam National University, Daejeon 34134, Republic of Korea}

\author{Yeonsang Park}
\email{yeonsang.park@cnu.ac.kr}
\thanks{Corresponding author}
\affiliation{Department of Physics, Chungnam National University, Daejeon 34134, Republic of Korea}

\date{\today}

\begin{abstract}
Long-context LLM inference is bottlenecked not by compute
but by the memory bandwidth required to scan the KV cache
at every decode step---a cost that grows linearly with context
length.
The semiconductor industry increasingly acknowledges this shift:
NVIDIA's Vera Rubin architecture dedicates an entire DPU (ICMS)
to KV cache management with flash-backed storage and
hardware-assisted prefetch---an architectural bet confirming that
memory, not arithmetic, is the first-class
constraint.

Recent photonic accelerators have demonstrated impressive
throughput for dense attention
computation.
However, these approaches inherit the same $O(n)$ memory scaling
as electronic attention when applied to long contexts.
We observe that the real leverage point is the coarse
block-selection step: a memory-bound similarity search that
determines which KV blocks to fetch.
We identify, for the first time, that this task is
\emph{structurally matched} to the photonic broadcast-and-weight
paradigm---the query fans out to all candidates via passive
splitting, signatures are quasi-static (matching electro-optic
MRR programming), and only rank order matters (relaxing precision
to 4--6 bits).
Crucially, the photonic advantage \emph{grows with context
length}: as $N$ increases, the electronic scan cost rises
linearly while the photonic evaluation remains $O(1)$.

We instantiate this insight in \prism{}, a thin-film lithium
niobate (TFLN) similarity engine.
Hardware-impaired needle-in-a-haystack evaluation on Qwen2.5-7B
confirms 100\% accuracy from 4K through 64K tokens at $k{=}32$,
with $32\times$ traffic reduction.
\prism{} achieves a four-order-of-magnitude energy advantage
over GPU baselines at practical context lengths ($n \geq 4$K).
\end{abstract}

\maketitle


\input{sections/introduction}           
\input{sections/background}             
\input{sections/architecture}           
\input{sections/hardware_analysis}      
\input{sections/system_evaluation}      
\input{sections/scaling_analysis}       
\input{sections/discussion}             
\input{sections/conclusion}             

\clearpage  


\input{sections/supplementary}

\bibliography{references}

\end{document}

%% file: sections/introduction.tex
\section{Introduction}
\label{sec:introduction}

The dominant cost of large language model (LLM) inference is no longer
floating-point arithmetic.
As autoregressive decoding generates one token at a time, each step
requires reading the full key--value (KV) cache accumulated over all
previous tokens, computing attention scores, and writing the result
back.
For a model with $L$ layers and $H$ attention heads, each storing
key and value vectors of dimension $d_h$, the KV cache occupies
$2LHd_h$ bytes per token (at half precision), growing linearly with
context length~$n$.
At $n = \num{128000}$ tokens, a 70-billion-parameter model's KV cache
can exceed \SI{40}{\giga\byte}---comparable to the entire model weight
footprint---and the memory bandwidth required to stream this cache
at every decode step far exceeds the compute throughput of modern
GPUs~\cite{FlashAttention2023}.

This memory wall is intensifying~\cite{Gholami2024MemWall}.
Context windows are expanding aggressively: GPT-4~\cite{OpenAI2023GPT4} and
Gemini~\cite{GeminiTeam2024} pushed context to 128K tokens,
Llama~3.1 supports
\num{128000} tokens~\cite{Llama3_2024}, Qwen2.5 extends to one
million~\cite{Yang2025Qwen1M}, and multi-agent and retrieval-augmented
generation (RAG) workloads routinely concatenate documents into
contexts of hundreds of thousands of tokens.
NVIDIA's response in its Vera Rubin architecture is telling: the
\emph{Intelligent Connectivity and Memory Switch} (ICMS), built on
the BlueField-4 data processing unit (DPU), adds a flash-based
KV cache tier that can hold terabytes of context, together with
hardware-assisted eviction and prefetch logic~\cite{NVIDIA_VeraRubin2026,
NVIDIA_ICMS2026}.
This architectural bet confirms that KV cache management is now a
first-class system design problem.

Photonic circuits on thin-film lithium niobate (TFLN) offer a set of physical properties that are
uniquely matched to this bottleneck.
A wavelength-division-multiplexed (WDM) laser comb encodes a
$d$-dimensional vector onto $d$ co-propagating wavelengths in a
single waveguide; a $1 \times N$ passive splitter then
\emph{broadcasts} identical copies of that vector to $N$ output
channels with no additional energy cost beyond splitting loss.
At each channel, a bank of microring resonators (MRRs)---one per
wavelength---applies programmable transmission weights, and a
broadband photodetector integrates over all wavelengths, yielding
the analog inner product in a single optical transit
($\sim$\SI{10}{\pico\second} per \si{\milli\meter}).
The entire $d \times N$ matrix--vector product thus completes in
$O(1)$ latency, with energy scaling dominated by weight-programming overhead
rather than memory-access energy.
This \emph{broadcast-and-weight} paradigm~\cite{Tait2014,Tait2016}
converts the memory-bandwidth-bound electronic problem into an
optically parallel computation.

Existing demonstrations of photonic neural-network accelerators have
focused almost exclusively on dense matrix--vector multiplication
for inference in convolutional and fully connected
networks~\cite{Hua2025Photonic,Zhu2024LighteningTransformer,Tian2025PTC,Fu2024ONN}.
In particular, Tian et al.\ demonstrated a photonic transformer chip
(PTC) that implements full attention via coherent optical interference
with runtime-programmable Mach--Zehnder meshes~\cite{Tian2025PTC};
however, full-attention photonic computation faces the same $O(n)$
memory scaling as electronic attention when applied to long contexts.
In contrast, the coarse block-selection step in KV cache retrieval is
\emph{not} a dense neural-network layer---it is a large-scale
\emph{similarity search}: a single query vector must be compared
against $N \sim 10^3$--$10^4$ stored signatures, and only the
top-$k$ matches are needed.
This search problem has three properties that make it a more
natural fit for the broadcast-and-weight architecture than
general-purpose matrix multiplication:
(i)~the query is broadcast identically to all channels, perfectly
matching the optical fan-out;
(ii)~the stored signatures are quasi-static (updated every
64--512 tokens), so MRR weights can be programmed via
electro-optic tuning (Pockels effect); and
(iii)~only rank order matters, relaxing the precision requirement to
4--6 bits.
We therefore propose the concept of a \emph{photonic broadcast
search}---an application of photonic broadcast-and-weight hardware
not as a general neural-network accelerator, but as a specialized
similarity engine for memory-intensive search tasks.

A crucial observation simplifies the problem.
Not all attention heads actually need the full cache.
Recent work on \emph{retrieval heads}~\cite{Wu2025RetrievalHead,
DuoAttention2024, RazorAttention2024} has shown that attention heads
split into two categories: \emph{retrieval} heads that attend to
tokens far from the current position, and \emph{streaming} heads
that attend primarily to nearby tokens and ``attention sinks.''
The fraction classified as retrieval heads is threshold-dependent:
DuoAttention identifies approximately 25\% of heads as retrieval
heads in MHA models and approximately 50\% in GQA models via
learned gating optimization~\cite{DuoAttention2024}, while our
profiling on Qwen2.5-7B finds over 90\% at a relaxed threshold
($\tau{=}0.3$; \cref{sec:system_eval}).
This discrepancy reflects differing identification criteria rather
than a contradiction---the key insight is that only the retrieval
subset requires distant block fetches.

This asymmetry has motivated a family of \emph{block-level selection}
methods that implement a coarse candidate selection step followed
by fine attention over only the selected blocks on the
GPU~\cite{Quest2024,RocketKV2024,InfLLM2024,Liu2025ChunkKV,Li2025KVSurvey}.
Complementary strategies include token-level eviction
(H\textsubscript{2}O~\cite{H2O2024}, StreamingLLM~\cite{StreamingLLM2023,Hooper2024KVQuant}),
two-stage coarse--fine retrieval (RocketKV~\cite{RocketKV2024}), and
hardware-assisted caching (NVIDIA ICMS~\cite{NVIDIA_ICMS2026}).
All electronic approaches share a common limitation: the coarse
selection step itself consumes memory bandwidth proportional to
the number of stored blocks.
Recent analysis confirms that this block selection phase can
consume the majority of total KV retrieval
latency~\cite{Liu2026FreeKV}.
A photonic inner-product engine can break this scaling by performing
all $N$ similarity evaluations in parallel, using wavelength
multiplexing to avoid the sequential memory access pattern entirely.

We propose \prism{} (\textbf{P}hotonic \textbf{R}anking via
\textbf{I}nner-product \textbf{S}imilarity with \textbf{M}icroring
weights), a TFLN photonic similarity engine that realizes the
photonic broadcast search concept for KV cache block selection.
\prism{} encodes the query sketch onto $d$ WDM wavelength channels,
broadcasts it to $N$ parallel MRR weight-bank channels via a
$1 \times N$ optical splitter, and computes all $N$ similarity
scores---each as an analog optical dot product
$I_n \propto \sum_{j} w_{n,j}\,s_j$---in $O(1)$ optical latency.
A compact electronic top-$k$ comparator selects the highest-scoring
block indices, and only the corresponding KV blocks are fetched
from memory.

\begin{figure*}[!tp]
  \centering
  \includegraphics[width=\textwidth]{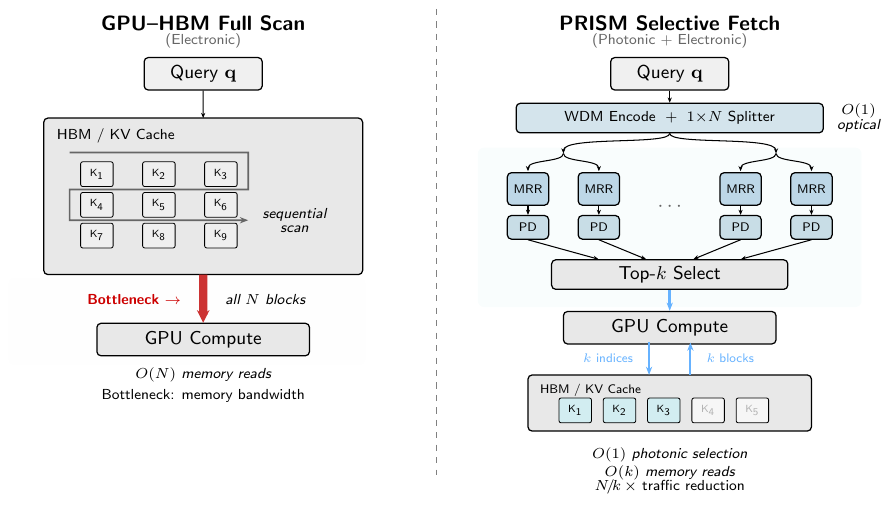}
  \caption{%
    Conceptual comparison of KV cache access strategies.
    \textbf{Left:} Electronic GPU full scan---the processor
    sequentially reads all $N$ KV blocks from HBM to compute
    attention, bottlenecked by memory bandwidth.
    \textbf{Right:} \prism{} photonic block selection---the query
    is broadcast optically to all $N$ signature channels in parallel;
    only the top-$k$ highest-scoring blocks are fetched from memory,
    reducing traffic by $N/k$ times.
  }
  \label{fig:concept_comparison}
\end{figure*}

\Cref{fig:concept_comparison} contrasts the conventional electronic
full-scan approach with the \prism{} photonic block-selection pipeline.

Our contributions are as follows:
\begin{enumerate}[leftmargin=*,itemsep=2pt]
  \item \textbf{Photonic broadcast search architecture.}
    We propose and analyze a photonic similarity engine
    based on the broadcast-and-weight paradigm, specifically
    designed for the KV cache block-selection task.
    We present a complete optical power budget analysis
    covering splitting loss, MRR insertion loss, and photodetector
    noise floors, and derive the signal-to-noise ratio (SNR)
    requirements for reliable top-$k$ ranking
    (\cref{sec:architecture} and \cref{sec:hardware}).
  \item \textbf{Hardware-aware impairment modeling and NIAH validation.}
    We build a device-level impairment model incorporating weight
    quantization (4--8 bits), residual thermal drift, insertion loss chains,
    photodetector noise, and MRR crosstalk, and show that recall
    degrades by less than 10\% under realistic conditions.
    End-to-end needle-in-a-haystack (NIAH) evaluation with
    Qwen2.5-7B demonstrates that MRR-selected block-sparse
    attention matches full-attention accuracy at context lengths
    from 4K to \textbf{64K tokens} (within the model's native
    context window), while replacing the electronic
    selection with photonic $O(1)$-latency computation.
    Beyond 64K, model-intrinsic accuracy degrades independent
    of block selection (\cref{sec:hardware}).
  \item \textbf{Photonic scaling analysis.}
    We derive energy and latency models for \prism{} and electronic
    baselines (GPU full scan, GPU ANN, NVIDIA ICMS), identifying
    the context-length crossover point above which \prism{} is
    favorable, and analyze how the photonic architecture scales
    to million-token contexts (\cref{sec:scaling}).
  \item \textbf{Retrieval head analysis and signature design.}
    We systematically profile retrieval-head ratios across
    Qwen2.5-7B and Qwen3-8B, confirming that
    over 90\% of KV heads are retrieval heads (at threshold $\tau = 0.3$),
    and evaluate block-level signatures, demonstrating that mean-key
    projection achieves 77.3\% recall@8 with $d = 32$
    (\cref{sec:system_eval}).
\end{enumerate}

%% file: sections/background.tex
\section{Background}
\label{sec:background}

\subsection{KV Cache in Transformer Inference}
\label{sec:bg_kvcache}

The core of modern LLMs is the multi-head self-attention
mechanism~\cite{Vaswani2017}.
Given an input sequence of $n$ tokens embedded as
$\mat{X} \in \mathbb{R}^{n \times d_{\text{model}}}$, each attention
head $h$ in layer $\ell$ projects the input into queries, keys, and
values:
\begin{align}
  \mat{Q}^{(\ell,h)} &= \mat{X}\,\mat{W}_Q^{(\ell,h)}, \nonumber\\
  \mat{K}^{(\ell,h)} &= \mat{X}\,\mat{W}_K^{(\ell,h)}, \nonumber\\
  \mat{V}^{(\ell,h)} &= \mat{X}\,\mat{W}_V^{(\ell,h)},
  \label{eq:qkv}
\end{align}
where $\mat{W}_Q, \mat{W}_K, \mat{W}_V \in \mathbb{R}^{d_{\text{model}}
\times d_h}$ and $d_h = d_{\text{model}} / H$ is the per-head
dimension.
The attention output is computed as
\begin{equation}
  \text{Attn}(\mat{Q}, \mat{K}, \mat{V})
  = \text{softmax}\!\left(\frac{\mat{Q}\,\mat{K}^T}{\sqrt{d_h}}\right)
    \mat{V}.
  \label{eq:attention}
\end{equation}

During the autoregressive \emph{decode phase}, the model generates
one token at a time.
At step $t$, only the new query vector
$\vect{q}_t \in \mathbb{R}^{d_h}$ is computed, but the attention
score requires the inner product of $\vect{q}_t$ with all $t$
previously cached key vectors:
\begin{equation}
  \alpha_{t,i} = \frac{\vect{q}_t \cdot \vect{k}_i}{\sqrt{d_h}},
  \quad i = 1, \ldots, t.
  \label{eq:decode_attn}
\end{equation}
The KV cache stores $\mat{K}^{(\ell,h)}$ and $\mat{V}^{(\ell,h)}$
for all layers and heads, consuming memory
\begin{equation}
  M_{\text{KV}} = 2\,L\,H_{\text{KV}}\,d_h\,n\,b_{\text{prec}},
  \label{eq:kv_memory}
\end{equation}
where $H_{\text{KV}}$ is the number of KV heads (which equals $H$
for multi-head attention but is reduced under grouped-query
attention, GQA~\cite{Ainslie2023GQA,Shazeer2019MQA}) and $b_{\text{prec}}$ is the byte width per element
(2 for BF16).
For Llama-3.1-8B ($L{=}32$, $H_{\text{KV}}{=}8$ with 4-group GQA,
$d_h{=}128$) at $n = \num{128000}$,
\cref{eq:kv_memory} gives $M_{\text{KV}} \approx \SI{16}{\giga\byte}$,
which already consumes a substantial fraction of GPU HBM and grows
linearly with $n$.

Crucially, the decode phase is \emph{memory-bandwidth-bound}:
each generated token requires reading the entire KV cache but
performs only $O(n \cdot d_h)$ multiply-accumulate operations
per head.
The arithmetic intensity (FLOPs per byte) is $1/(2d_h) \ll 1$,
far below the compute-to-bandwidth ratio of modern GPUs
(\SIrange{50}{200}{FLOP/B}), leaving the compute units idle while
waiting for data~\cite{FlashAttention2023}.

\subsection{Retrieval Heads and Selective Attention}
\label{sec:bg_retrieval}

The observation that not all attention heads require the full KV
cache was formalized by DuoAttention~\cite{DuoAttention2024} and
RazorAttention~\cite{RazorAttention2024}.
These works define a \emph{retrieval ratio} $R_h^{(\ell,h)}$ for
each head as the fraction of attention mass that falls outside a
local window of size $w$:
\begin{equation}
  R_h^{(\ell,h)} = 1 - \frac{1}{T}\sum_{t=1}^{T}
  \sum_{i=\max(1, t-w)}^{t} \alpha_{t,i}^{(\ell,h)},
  \label{eq:retrieval_ratio}
\end{equation}
where $\alpha_{t,i}^{(\ell,h)}$ is the attention weight from
\cref{eq:decode_attn} and $T$ is the total sequence length of a
calibration corpus.
Heads with $R_h > \tau$ (typically $\tau \approx 0.1$) are classified
as \emph{retrieval heads}; the rest are \emph{streaming heads}.

Empirically, DuoAttention identifies approximately 25\% (MHA) to
50\% (GQA) of heads as retrieval heads via learned gating
optimization~\cite{DuoAttention2024}.
Streaming heads can be served with a small sliding-window cache
(e.g., $w = 256$), drastically reducing their memory footprint.
However, retrieval heads still require access to the full context,
making their KV traffic the dominant bottleneck.

\subsection{Photonic Similarity Engine}
\label{sec:bg_photonic}

As noted in \cref{sec:introduction}, the coarse block-selection step
is a similarity search whose properties---identical query fan-out,
quasi-static weights, and rank-order-only output---make it a natural
fit for photonic broadcast-and-weight hardware.
We now review the key photonic concepts underlying this match.

\paragraph{Broadcast-and-weight architecture.}
Tait \textit{et al.}~\cite{Tait2014,Tait2016} introduced the
\emph{broadcast-and-weight} (B\&W) paradigm for neuromorphic
photonic networks.
In this architecture, $d$ input signals are encoded on distinct
wavelengths $\lambda_1, \ldots, \lambda_d$ and broadcast via a
$1 \times N$ optical splitter to $N$ output channels.
Each output channel contains $d$ microring resonators (MRRs),
each tuned to one wavelength, whose transmission coefficients
serve as programmable weights $w_{n,j}$ for channel $n$ and
wavelength $j$.
A wavelength-insensitive photodetector at each output integrates
over all wavelengths, yielding the photocurrent:
\begin{equation}
  I_n = \mathcal{R}\,P_0 \sum_{j=1}^{d} w_{n,j}\,s_j,
  \label{eq:bw_inner_product}
\end{equation}
where $\mathcal{R}$ is the detector responsivity, $P_0$ the
per-channel optical power after splitting, and $s_j$ the query
signal on wavelength $\lambda_j$.
The photocurrent $I_n$ is thus proportional to the inner product
$\vect{w}_n \cdot \vect{s}$---precisely the similarity score
between stored signature $n$ and the broadcast query.
This operation completes in a single optical transit time
($\sim$\SI{10}{\pico\second} per \si{\milli\meter}), independent
of $d$ and $N$ (up to splitting loss limits).

\paragraph{WDM spectral encoding.}
The query vector is encoded in the \emph{spectral domain}: each
component $s_j$ modulates the optical power on a dedicated
wavelength channel $\lambda_j$, so the full $d$-dimensional vector
propagates as a single multi-wavelength beam in one waveguide.
This spectral encoding is distinct from \emph{spatial encoding},
where each component occupies a separate waveguide, because it
enables the key broadcast step---splitting one waveguide into $N$
copies---with no additional multiplexing hardware.
Channel spacings of \SIrange{0.8}{1.6}{\nano\meter} within the
C-band support $d = 32$--$128$ channels using standard dense WDM
(DWDM) laser combs and MRR filter banks.

\paragraph{Comparison with other photonic paradigms.}
Alternative photonic architectures---MZI meshes~\cite{Reck1994,Clements2016,Shen2017,Fu2024ONN}
and coherent processors---require $O(d^2)$ elements or global phase
stability, and do not naturally support the one-to-many fan-out
needed for similarity search.
The broadcast-and-weight paradigm uses incoherent intensity-domain
processing, where each MRR operates independently and the
photodetector sums power rather than field amplitude, eliminating
the need for global phase coherence and making it uniquely suited
to the block-selection task.

\paragraph{MRR weight banks.}
Each output channel employs $d$ microring resonators whose
electro-optically tunable transmission implements programmable
weights $w \in [0,1]$.
The MRR physics and TFLN-specific device parameters are detailed
in Secs.~\ref{sec:arch_weightbank} and~\ref{sec:arch_device}.

\paragraph{WDM-based matrix--vector multiplication.}
Scalable MRR weight banks with up to 16 wavelength channels and
${\sim}$7-bit precision have been
demonstrated~\cite{Xu2005,Zhou2022,Huang2020}, and recent
large-scale photonic accelerators validate integration beyond
\num{16000} components~\cite{Hua2025Photonic,Zhu2024LighteningTransformer,Fu2024ONN,Zhang2025PNN}.

The key advantage of this photonic approach for the KV cache
selection problem is that the ``weight matrix''---the collection
of block signatures---is quasi-static and can be programmed into
MRR resonances via electro-optic tuning, while the ``input vector''---the
query sketch---changes at every decode step but is broadcast
optically to all $N$ channels simultaneously.
This decoupling of weight programming rate from inference rate
is what enables the $O(1)$ latency scaling that electronic
approaches cannot match.
While Lightening-Transformer~\cite{Zhu2024LighteningTransformer}
targets full attention computation, \prism{} takes a complementary
approach: accelerating only the lightweight block-selection ranking
task, which requires lower precision and fewer channels, making
the photonic implementation more practical.

%% file: sections/architecture.tex
\section{Photonic Retrieval Architecture}
\label{sec:photonic_retrieval_architecture}
\label{sec:architecture}  

\subsection{System Overview}
\label{sec:arch_overview}

\prism{} is a photonic similarity engine that sits between the KV cache
storage (HBM or flash-backed ICMS) and the GPU's attention compute
units.
It does not replace any part of the GPU pipeline; rather, it acts
as a \emph{photonic broadcast search} module that determines which
KV cache blocks should be fetched for each retrieval head at each
decode step.

\begin{figure*}[!tp]
  \centering
  \includegraphics[width=\textwidth,height=0.78\textheight,keepaspectratio]{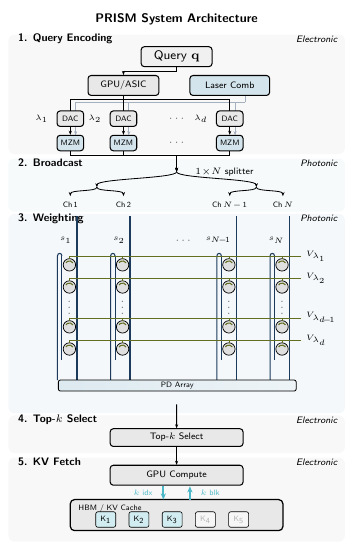}
  \caption{%
    \prism{} system architecture (five-stage pipeline).
    \textbf{Stage~1 (Query Encoding):}
    The GPU/ASIC computes the query sketch $\vect{q} = [q_1, \ldots, q_d]$
    and encodes each component onto a WDM wavelength via DAC-driven
    modulators, producing a WDM query signal where $P(\lambda_j) = q_j$.
    \textbf{Stage~2 (Broadcast):}
    A $1 \times N$ optical splitter distributes identical copies of
    the $d$-wavelength signal to all $N$ signature channels
    (splitting loss: $-10\log_{10}N$\,dB).
    \textbf{Stage~3 (Signature Weighting):}
    Each channel passes through a row of $d$ MRRs on the TFLN photonic chip;
    the transmission $t_{ij} = s_{ij}$ of each MRR is electro-optically
    programmed via DC bias electrodes to encode the block signature weight,
    performing wavelength-selective multiplication
    $P_{\text{out}}(\lambda_j) = q_j \times s_{ij}$.
    \textbf{Stage~4 (Summation):}
    Broadband photodetectors integrate all wavelengths, yielding
    photocurrents $I_i = \mathcal{R} \cdot \sum_j (q_j \cdot s_{ij})$
    that are proportional to the inner product $\vect{q} \cdot \vect{s}_i$.
    \textbf{Stage~5 (Top-$k$ Selection):}
    ADCs digitize the $N$ photocurrents, a digital top-$k$ selector
    identifies the $k$ highest-scoring block indices, and a memory
    controller fetches only those KV blocks from HBM/flash storage.
  }
  \label{fig:system_overview}
\end{figure*}

The system operates as a five-stage pipeline, illustrated in
\cref{fig:system_overview}.
For each retrieval head at each decode step:
\begin{enumerate}[leftmargin=*,itemsep=2pt]
  \item The GPU computes the query vector $\vect{q}_t$ and applies
        the signature projection to obtain a $d$-dimensional query
        sketch $\vect{s}_q$.
  \item $\vect{s}_q$ is converted to the optical domain and broadcast.
  \item The photonic weight bank computes $N$ inner products in
        parallel.
  \item Photodetectors produce $N$ analog similarity scores.
  \item A digital top-$k$ selector identifies the best blocks, and
        only those blocks are fetched from KV cache storage.
\end{enumerate}
The GPU then computes exact attention over the selected blocks plus
the local sliding window, producing the final attention output.

\subsection{Signature Encoding}
\label{sec:arch_signature}

The performance of \prism{} depends critically on the quality of the
block-level signatures programmed into the MRR weight banks.
Since signature encoding defines the input interface between the
digital LLM pipeline and the photonic engine, we describe it first.
We consider four signature construction methods.

\paragraph{Mean key.}
The simplest approach averages the key vectors within each block:
\begin{equation}
  \vect{\sigma}_n = \frac{1}{B}\sum_{i \in \text{block}\,n}
  \vect{k}_i^{(\ell,h)}.
  \label{eq:sig_mean}
\end{equation}
This preserves the original key-space geometry but requires
$d_h$-dimensional signatures (e.g., $d_h = 128$), demanding a
correspondingly large number of MRRs per channel.

\paragraph{PCA projection.}
Principal component analysis over the key distribution yields a
projection matrix $\mat{P} \in \mathbb{R}^{d \times d_h}$
($d \ll d_h$) that captures the dominant variance directions.
The signature becomes $\vect{\sigma}_n = \mat{P}\,\bar{\vect{k}}_n$,
reducing the MRR count per channel from $d_h$ to $d$.

\paragraph{Random projection.}
The Johnson--Lindenstrauss (JL) lemma guarantees that a random
Gaussian matrix $\mat{R} \in \mathbb{R}^{d \times d_h}$ with
$d = O(\epsilon^{-2} \log N)$ preserves pairwise distances (and
hence inner-product rankings) to within a factor $1 \pm \epsilon$
with high probability~\cite{JohnsonLindenstrauss1984}.
The query sketch is computed identically:
$\vect{s}_q = \mat{R}\,\vect{q}_t$.
Random projection is attractive because it requires no training
and provides worst-case guarantees.

\paragraph{Learned projection.}
A trainable linear layer $\mat{W}_{\text{proj}} \in \mathbb{R}^{d
\times d_h}$ is optimized end-to-end to maximize recall@$k$
on a calibration set.
This can outperform random projections when the key distribution
has exploitable structure, but requires per-model training.

\paragraph{Balanced photodetection.}
The add-drop MRR configuration provides both through-port and drop-port
outputs simultaneously.  A balanced photodetector pair measures the
differential photocurrent
$I_{\text{bal}} = I_{\text{through}} - I_{\text{drop}}$,
yielding a signed weight
$w_{n,j} = T_{\text{through}}(\lambda_j) - T_{\text{drop}}(\lambda_j) \in [-1,\,+1]$.
On-resonance (minimum through-port transmission), $w \approx -1$;
fully detuned, $w \approx +1$.
This eliminates the need for split encoding or ReLU projection,
enabling direct signed inner products with $d$~MRRs per channel
(half the count of split encoding) while preserving full sign information.

\subsection{WDM Query Broadcast}
\label{sec:arch_broadcast}
\label{sec:arch_stages}  

The $d$-dimensional query sketch $\vect{s}_q = [s_1, s_2, \ldots, s_d]$
is converted from the digital domain by $d$ digital-to-analog converters
(DACs), each driving a Mach--Zehnder modulator (MZM)~\cite{Wang2018TFLN} that impresses
the value $s_i$ onto wavelength $\lambda_i$ from a WDM laser comb source.
The modulated signals are multiplexed into a single waveguide carrying
$d$ wavelength-encoded values~\cite{Shastri2021,Totovic2022WDM}.

The DAC resolution requirement is modest: since the task is ranking
rather than exact computation, 4--6 bits of input precision suffice
(\cref{sec:hw_impairments}).
This relaxation is critical because high-resolution, high-speed DACs
are a major energy cost in photonic accelerators.
At 4-bit resolution, a DAC operating at \SI{1}{\giga Sa/s} consumes
approximately \SI{0.5}{\milli\watt}~per channel.

The multiplexed $d$-wavelength signal is then split into $N$ copies by a
$1 \times N$ optical splitter tree.
Each copy carries the full query sketch, attenuated by the splitting
loss:
\begin{equation}
  L_{\text{split}} = 10\log_{10}(N) + \alpha_{\text{excess}}\,
  \lceil\log_2 N\rceil \quad [\text{dB}],
  \label{eq:split_loss}
\end{equation}
where $\alpha_{\text{excess}} \approx \SI{0.2}{\deci\bel}$ per stage
for optimized $1 \times 2$ directional couplers.
For $N = 1024$ blocks, the total splitting loss is approximately
\SI{32}{\deci\bel}, requiring a laser source power of
\SIrange{10}{20}{\deci\bel m} to maintain adequate signal-to-noise
ratio (SNR) at the photodetectors.

To manage loss, the $N$ channels can be organized into $N_{\text{bank}}$
independent banks, each serving $N / N_{\text{bank}}$ channels with a
separate splitter tree.
This reduces per-bank splitting loss at the cost of additional laser
sources or optical amplifiers.
The key point is that the broadcast is \emph{passive and energy-free}:
the same query vector reaches all $N$ channels simultaneously, with
no per-channel memory access or data movement.

\subsection{MRR Weight Bank Similarity Engine}
\label{sec:arch_weightbank}

Each of the $N$ output channels contains a linear array of $d$ MRRs,
one per wavelength channel.
The $j$-th MRR in channel $n$ is electro-optically tuned so that its
transmission at wavelength $\lambda_j$ encodes the signature weight
$w_{n,j}$:
\begin{equation}
  \begin{split}
    P_{\text{out},n}(\lambda_j) &= w_{n,j} \cdot P_{\text{in}}(\lambda_j), \\
    w_{n,j} &= T_{\text{through},n}(\lambda_j) - T_{\text{drop},n}(\lambda_j) \in [-1,\,+1].
  \end{split}
  \label{eq:mrr_weight}
\end{equation}
The total number of MRRs in the system is $d \times N$.
For $d = 64$ and $N = 1024$, this yields \num{65536} MRRs---a
large but feasible integration scale for current photonic
platforms~\cite{Huang2020}.

Weight programming occurs at the block completion rate.
When a new KV cache block of $B$ tokens is completed, the
corresponding column of MRR weights is updated via electro-optic
(Pockels) tuning with sub-nanosecond response time.
During steady-state decoding, the weight bank is static and the
only dynamic signal is the broadcast query sketch.
Because TFLN EO tuning is capacitive, the MRR weight bank consumes
near-zero static power---only switching energy
(${\sim}\SI{5}{\femto\joule}$ per weight update) is required.
This \SI{5}{fJ} figure refers to the MRR electrode charging energy alone;
the total switching energy including CMOS driver circuits is estimated
at 50--500\,fJ.

Each channel terminates in a broadband (wavelength-insensitive)
photodetector~\cite{Lischke2021} that integrates the optical power across all $d$
wavelengths:
\begin{equation}
  \begin{split}
    I_n &= \mathcal{R} \sum_{j=1}^{d} \left[
      T_{\text{through},n}(\lambda_j) - T_{\text{drop},n}(\lambda_j)
    \right] s_j\,P_0 \\
    &= \mathcal{R} \sum_{j=1}^{d} w_{n,j}\,s_j\,P_0,
  \end{split}
  \label{eq:photodetection}
\end{equation}
where $w_{n,j} \in [-1,\,+1]$.
This is precisely an \emph{analog optical dot product}: the
photocurrent $I_n \propto \sum_{j=1}^{d} w_{n,j}\,s_j$ computes
the similarity score $\vect{w}_n \cdot \vect{s}_q$ between stored
block signature $n$ and the broadcast query~\cite{Peserico2023}, with no explicit
multiply-accumulate circuit.
The physics of broadband photodetection inherently performs the
summation---no electronic accumulator is needed~\cite{Feldmann2021}.

\subsection{Electronic Top-$k$ Interface}
\label{sec:arch_topk}

The $N$ photocurrents are converted to digital values by $N$ ADCs
and fed to a digital top-$k$ comparator network.
The comparator identifies the $k$ channels with the largest similarity
scores and outputs their indices.
For $k \ll N$, a partial-sort network suffices, with complexity
$O(N \log k)$ and latency of a few nanoseconds at
\SI{1}{\giga\hertz} clock.
The ADC resolution can be as low as 4--6 bits, since only the
rank ordering matters.

\subsection{Device Parameters}
\label{sec:arch_device}

\begin{figure}[!htb]
  \centering
  \includegraphics[width=\columnwidth]{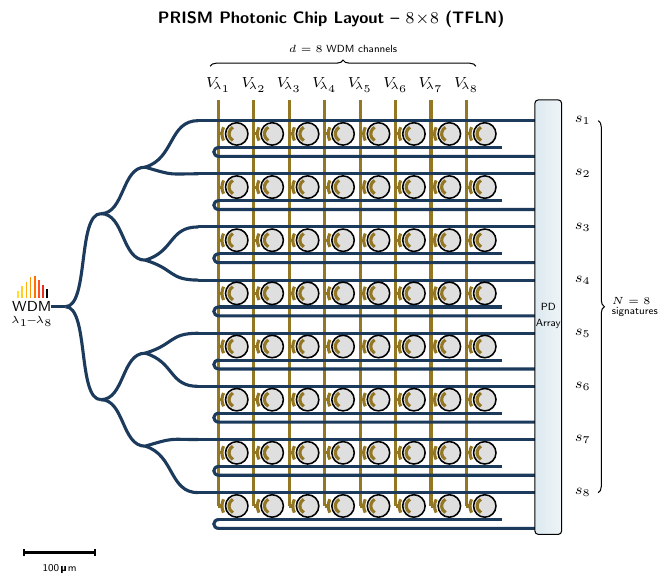}
  \caption{%
    \prism{} photonic chip layout for an $8 \times 8$ configuration
    ($d = 8$ WDM channels, $N = 8$ signature rows).
    Left: the WDM query input ($\lambda_1$--$\lambda_8$) enters and
    is split by cascaded $1 \times 2$ Y-junctions.
    Center: each row contains $d$ MRRs coupled to a bus waveguide
    with coupling gap of ${\sim}$200--300\,nm; EO DC bias electrodes
    program the MRR resonances to encode signature weights via the
    Pockels effect.
    Right: through-port and drop-port outputs route to balanced Ge-on-Si PD pairs
    (or optionally on-chip integrated photodetectors).
    Scale bar: 100\,\si{\micro\meter}.
    The layout scales to $d = 32$, $N = 256$ by increasing the
    splitter tree depth and the number of rows.
  }
  \label{fig:chip_layout}
\end{figure}

\begin{figure}[!htb]
  \centering
  \includegraphics[width=0.75\columnwidth]{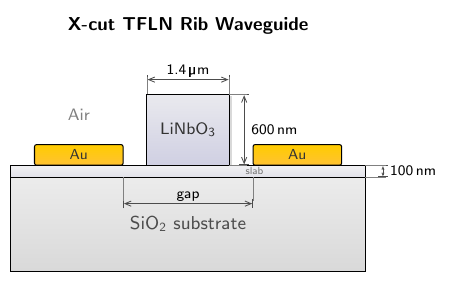}
  \caption{%
    X-cut TFLN rib waveguide cross-section.
    The rib is etched 500\,nm into a 600\,nm LN film on SiO$_2$,
    leaving a 100\,nm slab.
    Lateral Au electrodes apply DC bias for electro-optic (Pockels)
    tuning of the MRR resonance wavelength.
    Waveguide width: 1.4\,\si{\micro\meter}.
  }
  \label{fig:waveguide_crosssection}
\end{figure}

\Cref{tab:device_params} summarizes the assumed device parameters
for the thin-film lithium niobate (TFLN) photonic platform, based on
recent demonstrations of high-$Q$ TFLN micro-ring
resonators~\cite{Hu2025TFLN_NatComm,Zhu2024TFLN_HighQ} and
MRR weight bank architectures~\cite{Tait2017,Huang2020}.
The physical chip layout for an $8 \times 8$ demonstration
configuration is shown in \cref{fig:chip_layout}.

\begin{table}[!htb]
  \centering
  \caption{%
    \prism{} device parameters (TFLN platform).
  }
  \label{tab:device_params}
  \begin{tabular}{@{}lll@{}}
    \toprule
    Parameter & Value & Notes \\
    \midrule
    Platform & X-cut TFLN & 600\,nm LN/SiO$_2$ \\
    Waveguide & Rib, $1.4\!\times\!0.6\,\mu$m & 500\,nm etch \\
    MRR radius & $20\,\mu$m & FSR$\,\approx\,$8.3\,nm \\
    $Q_L$ & $\sim\!10^4$ & FDTD: 12{,}500 \\
    Extinction & $>$15\,dB & Add-drop \\
    Wt.\ precision & 5\,bit & EO resolution \\
    Tuning & EO (Pockels) & 28.5\,pm/V \\
    Static power & ${\sim}0$ & Capacitive EO \\
    Switch energy & ${\sim}$5\,fJ/ring & Per update \\
    Tuning speed & $<$1\,ns & EO response \\
    Modulator & TFLN MZM & $V_\pi L\!\sim\!$2\,V$\cdot$cm \\
    Photodetector & Balanced PD & Differential \\
    WDM ch. & $d\!=\!32$--128 & 1.6\,nm spacing$^{\dagger}$ \\
    Laser & Comb source & $\leq$100\,mW \\
    \bottomrule
  \end{tabular}

  \smallskip\noindent{\footnotesize $^{\dagger}$\,$d{=}32$--$64$ is realistic with current C+L band technology; $d{=}128$ requires C+L+S band operation and has not been experimentally demonstrated.}
\end{table}

The total MRR count ($d \times N$; \cref{eq:mrr_weight}) scales
with configuration as shown in \cref{tab:scaling_configs}.
Because TFLN electro-optic tuning is capacitive, the static power
consumption is near zero (\cref{sec:scale_thermal}).

%% file: sections/hardware_analysis.tex
\section{Photonic Hardware Analysis}
\label{sec:hardware}

We now incorporate realistic photonic device impairments into the
\prism{} simulation and quantify the optical link budget, noise
performance, and energy--latency tradeoffs against electronic
baselines.

\subsection{Device Impairment Modeling}
\label{sec:hw_impairments}

We model six impairment sources that degrade the ideal inner-product
computation of \cref{eq:bw_inner_product}~\cite{Bogaerts2012}:
(i)~weight quantization (4--8 bit DAC precision)~\cite{FerreiraeLima2022},
(ii)~thermal drift of MRR resonance wavelengths
($\sigma_{\text{drift}} = \SIrange{0.01}{0.1}{\nano\meter}$)~\cite{Padmaraju2014},
(iii)~MRR and waveguide insertion loss,
(iv)~photodetector shot and thermal noise
(NEP $\sim \SI{10}{\pico\watt/\sqrt{Hz}}$),
(v)~inter-channel MRR crosstalk (\SIrange{-15}{-30}{\deci\bel} isolation),
and (vi)~input DAC quantization noise.
\Cref{tab:impairment_params} summarizes the parameter ranges used
in the hardware simulation.
Full impairment models are provided in Supplementary Section~S1.

\subsection{Optical Link Budget}
\label{sec:hw_linkbudget}
\label{sec:hw_link_budget}

A critical question for any photonic accelerator is whether sufficient
optical signal-to-noise ratio (SNR) can be maintained across the
complete optical path~\cite{Shen2017}.
\Cref{tab:link_budget} traces the optical power from the laser source
to each photodetector for a representative configuration
($d = 32$, $N = 256$).

\begin{table*}[!htb]
  \centering
  \small
  \caption{%
    Optical link budget for $d = 32$, $N = 256$
    ($P_{\text{laser}} = \SI{20}{dBm}$).
  }
  \label{tab:link_budget}
  \begin{tabular}{@{}lrc@{}}
    \toprule
    Element & Loss (dB) & Cumulative (dBm) \\
    \midrule
    Laser output           & ---    & $+$20.0 \\
    Fiber--chip coupling   & $-$2.0 & $+$18.0 \\
    MZM modulator (avg.)   & $-$3.0 & $+$15.0 \\
    $1 \times 256$ splitter & $-$25.7$^a$ & $-$10.7 \\
    Waveguide (2\,cm)      & $-$1.0 & $-$11.7 \\
    $d{=}32$ MRR (balanced, worst-case drop) & $-$3.2$^b$ & $-$14.9 \\
    Chip--detector coupling & $-$1.0 & $-$15.9 \\
    \midrule
    \textbf{Per-PD optical power} & & $\mathbf{-15.9}$\,dBm \\
    \bottomrule
  \end{tabular}

  \vspace{2pt}
  {\footnotesize
    $^a$\,$10\log_{10}(256) + 0.2 \times 8 = 24.1 + 1.6 = 25.7$\,dB.\quad
    $^b$\,Drop-port 0.1\,dB + 31$\times$0.05\,dB through-port = 1.65\,dB;
    rounded to 3.2\,dB with alignment margin~\cite{Bogaerts2012}.}
\end{table*}

\paragraph{Balanced detection link budget.}
The link budget in \cref{tab:link_budget} traces the drop-port path
to the target photodetector.
In the balanced configuration used by \prism{}, each MRR channel
requires \emph{two} optical paths---through-port and
drop-port---each terminated by a separate photodetector and TIA.
The through-port path sees lower loss (no drop-port penalty),
so the drop-port budget above represents the worst case.
Consequently, balanced detection doubles the photodetector and TIA
count to $2N$ per wavelength channel; this overhead is reflected in
\cref{tab:prism_energy}.

At $P_{\text{PD}} = \SI{-15.9}{dBm} \approx \SI{25.7}{\micro\watt}$
per detector, the resulting photocurrent is
$I_{\text{ph}} = \mathcal{R} \cdot P_{\text{PD}} = 1.0 \times
\SI{25.7}{\micro\watt} = \SI{25.7}{\micro\ampere}$.
The electrical SNR at the detector is
\begin{equation}
  \text{SNR} = \frac{I_{\text{ph}}^2}{2eI_{\text{ph}}\Delta f
  + 4k_BT\Delta f / R_L + (\mathcal{R}\cdot\text{NEP})^2\Delta f},
  \label{eq:snr_link}
\end{equation}
where $\Delta f \approx \SI{1}{\giga\hertz}$ (matching the query
update rate) and $\text{NEP} = \SI{10}{\pico\watt/\sqrt{Hz}}$
(\cref{eq:supp_detector_noise}).
For $R_L = \SI{1}{\kilo\ohm}$ and $T = \SI{300}{\kelvin}$, we
obtain SNR~$\approx \SI{37.2}{dB}$---well above the minimum required
for reliable rank ordering~\cite{Sunny2021}.
(Note: $R_L = \SI{1}{\kilo\ohm}$ assumes a transimpedance amplifier
(TIA) front-end rather than \SI{50}{\ohm} termination.)

For larger bank sizes ($N = 1024$), the additional \SI{6}{dB}
splitting loss reduces the per-detector power to
$\SI{-21.9}{dBm} \approx \SI{6.5}{\micro\watt}$, yielding
SNR~$\approx \SI{25.5}{dB}$.
This remains adequate for top-$k$ ranking, as verified by the
recall analysis in \cref{sec:hw_recall}.
Beyond $N = 4096$ (SNR~$\approx \SI{13.5}{dB}$), the link budget
requires either a higher-power laser
($P_{\text{laser}} > \SI{26}{dBm}$) or the banked splitter
architecture described in \cref{sec:arch_stages}.

\begin{figure}[!tbp]
  \centering
  \includegraphics[width=\columnwidth,height=0.45\textheight,keepaspectratio]{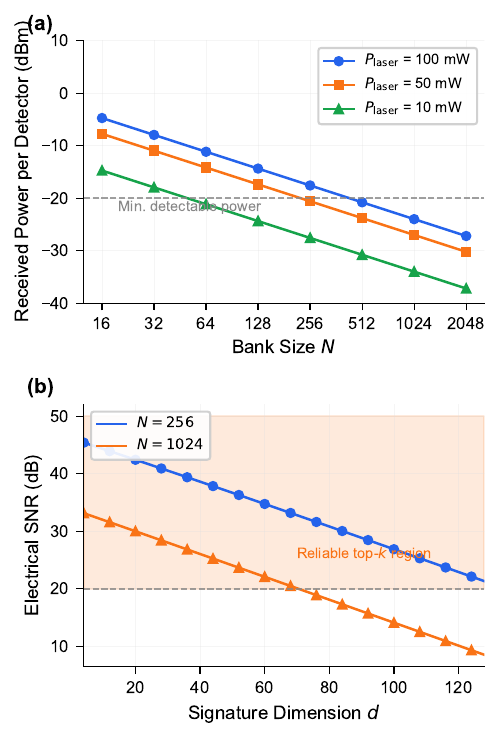}
  \caption{%
    Optical power budget analysis.
    (a)~Per-detector received power vs.\ bank size $N$ for three
    laser powers.
    The horizontal dashed line indicates the minimum detectable power
    ($\SI{-20}{dBm}$).
    (b)~Electrical SNR at the photodetector vs.\ signature dimension
    $d$ for $N = 256$ and $N = 1024$.
    The shaded region marks SNR~$> \SI{20}{dB}$, sufficient for
    reliable top-$k$ ranking.
  }
  \label{fig:power_budget}
\end{figure}

\Cref{fig:power_budget} illustrates the per-detector received power
and SNR as a function of the bank size $N$, clearly showing the
crossover point at which banked architectures or optical amplification
become necessary.

\begin{figure}[!htb]
  \centering
  \includegraphics[width=\columnwidth]{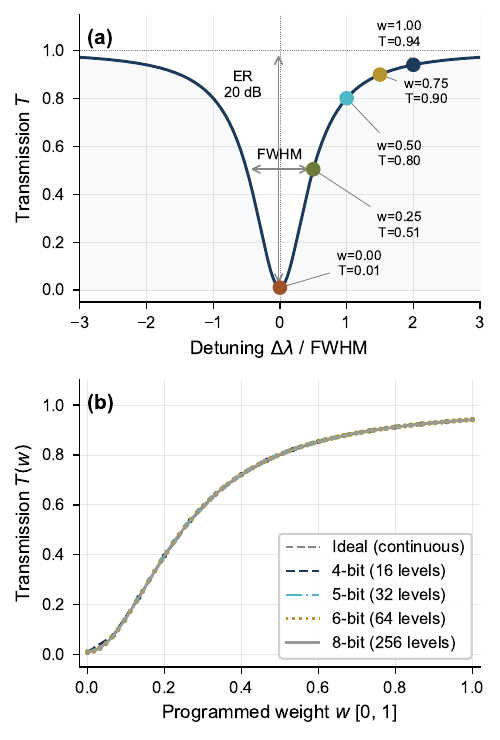}
  \caption{%
    MRR weight encoding principle.
    (a)~Through-port and drop-port transmission of a single add-drop MRR
    ($Q_L = 10{,}000$, ER~$= 20$\,dB).
    The balanced weight $w = T_{\text{through}} - T_{\text{drop}}$ maps from
    $-1$ (on-resonance) to $+1$ (fully detuned).
    (b)~Weight-to-balanced-transmission mapping for different DAC precisions.
  }
  \label{fig:mrr_lorentzian}
\end{figure}

\begin{table}[!htbp]
  \centering
  \small
  \caption{%
    Device impairment parameter ranges used in hardware simulation.
  }
  \label{tab:impairment_params}
  \begin{tabular}{@{}llc@{}}
    \toprule
    Impairment & Parameter & Range \\
    \midrule
    Weight quantization     & $b$ (bits)
      & 4--8 \\
    Thermal drift           & $\sigma_{\text{drift}}$ (\si{pm})
      & 10--100 \\
    MRR insertion loss      & IL$_{\text{MRR}}$ (\si{dB})
      & 0.02--0.05$^c$ \\
    Splitter excess loss    & $\alpha_{\text{excess}}$ (\si{dB/stage})
      & 0.1--0.3 \\
    Detector NEP            & (\si{pW/\sqrt{Hz}})
      & 1--20 \\
    MRR crosstalk           & Isolation (\si{dB})
      & $-$15 to $-$30 \\
    DAC resolution          & $b_{\text{DAC}}$ (bits)
      & 4--8 \\
    \bottomrule
  \end{tabular}

  \vspace{2pt}
  {\footnotesize $^c$\,Through-port IL per non-target MRR;
    drop-port (target MRR) IL is ${\sim}0.1$\,dB.}
\end{table}

\subsection{Recall Degradation Analysis}
\label{sec:hw_recall}

We inject impairments into the inner-product computation and measure
recall@$k$ degradation relative to the ideal (floating-point) baseline.
Individual impairment sweeps (quantization precision, thermal drift,
weight fidelity, and detector noise) are presented in
Supplementary Figs.~S1--S4.

\begin{figure*}[!tp]
  \centering
  \includegraphics[width=\textwidth]{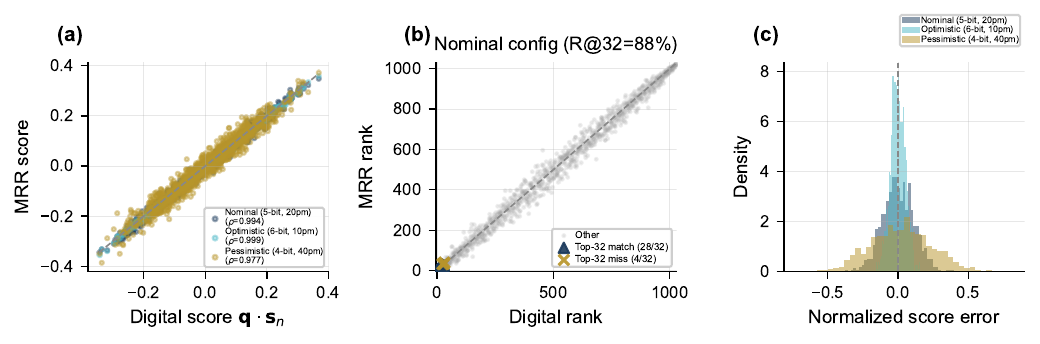}
  \caption{%
    Digital vs.\ MRR photonic inner-product comparison
    ($d = 32$, $N = 256$, $K = 8$).
    (a)~Score correlation between exact (FP64) and MRR-computed
    similarity for three hardware configurations.
    Pearson correlation $\rho > 0.98$ for all configs.
    (b)~Rank agreement for the nominal config (5-bit, \SI{20}{pm}):
    green triangles indicate correctly identified top-$K$ blocks
    (7/8 match, Recall@8~$= 88$\%).
    (c)~Normalised score error distributions;
    pessimistic config (4-bit, \SI{30}{pm}) shows wider tails but
    remains zero-centred.
  }
  \label{fig:digital_vs_photonic}
\end{figure*}

\begin{figure}[!htb]
  \centering
  \includegraphics[width=\columnwidth]{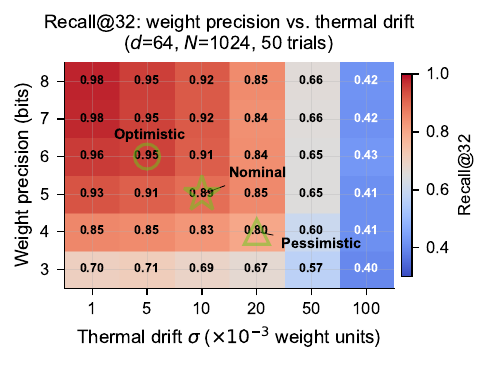}
  \caption{%
    Combined impairment sensitivity: Recall@8 as a function of weight
    precision (bits) and thermal drift $\sigma$ ($d = 32$, $N = 500$,
    50 Monte Carlo trials per cell).
    Markers indicate the three operating points studied in this work:
    nominal (5-bit, $\sigma = 0.01$), optimistic (6-bit, $\sigma = 0.005$),
    and pessimistic (4-bit, $\sigma = 0.02$).
    Recall exceeds 80\% for ${\geq}5$-bit precision and
    $\sigma \leq 0.02$.
  }
  \label{fig:combined_sensitivity}
\end{figure}

\paragraph{Combined impairments.}
We simulate the full impairment chain (quantization + drift +
loss + noise + crosstalk) using a Monte Carlo approach with
100 trials of 500 blocks ($d=32$).
\Cref{fig:digital_vs_photonic} visualises the effect for a single
trial: the MRR scores correlate strongly with the digital baseline
($\rho > 0.98$), and the top-$k$ ranking is largely preserved.
\Cref{fig:combined_sensitivity} maps the recall degradation as a
function of both weight precision and thermal drift magnitude,
identifying the operating region in which Recall@8 exceeds 80\%.

The combined recall degradation at $b = 6$, $\sigma_{\text{th}} = 0.01$,
and $\sigma_{\text{det}} = 0.01$ is approximately 8\%, yielding an
effective Recall@8 of $0.916 \pm 0.087$ (vs.\ 1.000 ideal).
Each impairment source individually contributes modestly
(5-bit quantization: 0.904, drift: 0.948, noise: 0.928), but their
combination remains above the 90\% threshold required for
effective block selection.

The recall degradation results establish the acceptable operating
region for the MRR weight bank.
End-to-end NIAH validation with MRR-simulated block selection,
confirming that these impairments do not degrade downstream task
accuracy, is presented in \cref{sec:niah_validation}.

\subsection{Energy Model}
\label{sec:hw_energy}

\Cref{tab:prism_energy} breaks down the energy per query evaluation
for the \prism{} system.
We define the energy metric as energy per inner-product evaluation
(i.e., per block scored per query per head).

\begin{table*}[!htb]
  \centering
  \small
  \caption{%
    \prism{} energy breakdown per query
    ($d = 64$, $N = 1024$, $k = 32$, TFLN platform).
  }
  \label{tab:prism_energy}
  \begin{tabular}{@{}lrc@{}}
    \toprule
    Component & Power (\si{mW}) & Energy/query (\si{pJ}) \\
    \midrule
    Laser source        & 20.0  & 180 \\
    TEC (thermal stab.)  & 1000  & 9000$^\dagger$ \\
    Voltage driver array & 5.0   & 45 \\
    DACs ($d$ channels) & 32.0  & 288 \\
    MZM modulators      & 6.4   & 58 \\
    EO bias (static)    & ${\sim}0$ & ${\sim}0$$^*$ \\
    Photodetectors ($2N$, balanced)      & 10.0   & 90 \\
    TIAs + ADCs ($2N$, balanced)         & 100.0  & 900 \\
    Top-$k$ logic       & 1.0   & 9 \\
    \midrule
    \textbf{Dynamic subtotal} & \textbf{174.4} & \textbf{1570} \\
    \textbf{Total (incl.\ TEC)} & \textbf{1174.4} & \textbf{10570} \\
    \bottomrule
  \end{tabular}

  \vspace{2pt}
  {\footnotesize $^*$TFLN EO tuning is capacitive;
    switching energy ${\sim}\SI{5}{fJ}$/ring.\\
    $^\dagger$TEC power is amortized across all heads and queries;
    per-query share $\ll \SI{1}{nJ}$ at realistic throughput.}
\end{table*}

A key advantage of the TFLN platform is the elimination of static
MRR tuning power.
TFLN electro-optic tuning via the Pockels effect is capacitive and
consumes near-zero static power (see \cref{sec:scale_thermal} for
the quantitative SOI comparison).
The only energy cost per weight update is the switching energy of
${\sim}\SI{5}{\femto\joule}$ per ring, which is negligible compared
to the dynamic optical and electronic components~\cite{Shastri2021}.
Note that while the total system power (${\sim}\SI{1.17}{W}$) is
dominated by the TEC, this is a fixed overhead shared across all
heads and queries; at typical decode throughput ($>$\num{1000}
tokens/s), the amortized TEC contribution per query is
$<\SI{1}{\micro\joule}$---still well below the GPU baseline.

For comparison, the H100 GPU full-scan baseline reads every KV
block signature once per query per head.
The energy per selection is
\begin{equation}
  \begin{split}
  E_{\text{scan}} &= 2\,d_h\,N\,b_{\text{prec}}\,E_{\text{byte}} \\
  &= 2 \times 128 \times 1024 \times 2 \times 31\;\text{pJ/B} \\
  &\approx \SI{16.3}{\micro\joule},
  \end{split}
  \label{eq:e_scan_gpu}
\end{equation}
where $d_h{=}128$ is the head dimension, $N{=}n/B{=}1024$ blocks
at 128K context ($B{=}128$), $b_{\text{prec}}{=}2$\,B (bf16), and
$E_{\text{byte}} \approx \SI{31}{pJ/B}$ ($\approx \SI{3.9}{pJ/bit}$,
standard HBM3 specification)~\cite{Choquette2023}.
Note that this baseline assumes GPU scans the full key dimension
$d_h{=}128$; if the GPU instead scans compressed $d{=}32$
signatures, the energy reduces to ${\sim}\SI{4.1}{\micro\joule}$
($4\times$ lower).
Even in this fairer comparison, \prism{}'s ${\sim}\SI{1570}{pJ}$
selection energy remains over three orders of magnitude below the
GPU scan~\cite{Feldmann2021}.
GPU ANN (FAISS IVF-PQ) reduces the full-key scan to
${\sim}$\SI{5}{\micro\joule} by scanning $O(\sqrt{N})$ centroids.
NVIDIA ICMS consumes ${\sim}$\SI{10}{\micro\joule}, estimated by
replacing $\text{BW}_{\text{HBM}}$ with the DPU's internal LPDDR5
bandwidth (${\sim}\SI{100}{GB/s}$) and assuming a similar scan
pattern over the flash-backed KV index.

\subsection{Latency Model}
\label{sec:hw_latency}

The \prism{} latency is the sum of the five pipeline stages:
\begin{equation}
  t_{\text{PRISM}} = t_{\text{DAC}} + t_{\text{opt}} +
  t_{\text{PD}} + t_{\text{ADC}} + t_{\text{top-}k},
  \label{eq:latency_prism}
\end{equation}
where the optical propagation time $t_{\text{opt}}$ includes the
modulator response, waveguide transit, and MRR ring-down time.

\begin{table}[!htb]
  \centering
  \small
  \caption{%
    \prism{} latency breakdown.
  }
  \label{tab:prism_latency}
  \begin{tabular}{@{}lrl@{}}
    \toprule
    Stage & Latency & Notes \\
    \midrule
    DAC & ${\sim}$1\,ns & 4-bit \\
    MZM & ${\sim}$0.1\,ns & Si depl.\ \\
    Opt.\ prop. & ${\sim}$0.5\,ns & 5\,cm \\
    MRR decay & ${\sim}$0.1\,ns & $Q{=}10^4$ \\
    PD & ${\sim}$0.2\,ns & Ge \\
    TIA+ADC & ${\sim}$2\,ns & 6-bit flash \\
    Top-$k$ & ${\sim}$5\,ns & CMOS \\
    \midrule
    \textbf{Total} & \textbf{${\sim}$9\,ns} & \\
    \bottomrule
  \end{tabular}
\end{table}

The total \prism{} latency of $\sim$\SI{9}{\nano\second} compares
favorably with the electronic baselines: GPU full scan
${\sim}$\SI{5}{\micro\second}, GPU ANN ${\sim}$\SI{1}{\micro\second},
and NVIDIA ICMS ${\sim}$\SI{0.5}{\micro\second}---representing a
${\sim}500\times$ speedup over full scan.
However, this comparison must account for the additional latency of
fetching the selected KV blocks from memory after \prism{} selection,
which adds \SIrange{0.5}{2}{\micro\second} depending on the memory
tier (HBM vs.\ flash).
The net latency benefit of \prism{} is therefore most pronounced when
the selection ratio $k/N$ is small and the KV cache resides in a
slow memory tier (e.g., flash in ICMS).
The crossover analysis quantifying these trade-offs across context
lengths and baselines is presented in \cref{sec:eval_crossover}.

\begin{figure}[!htb]
  \centering
  \includegraphics[width=\columnwidth]{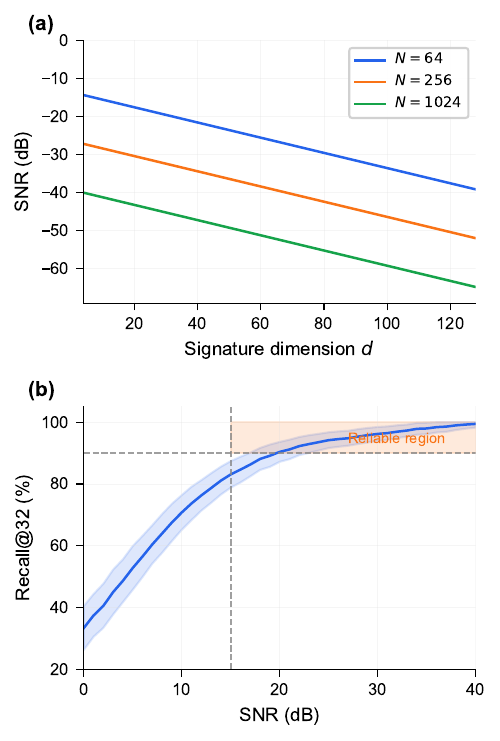}
  \caption{%
    SNR and recall analysis.
    (a)~Electrical SNR at the photodetector as a function of
    signature dimension $d$ for three bank sizes.
    (b)~Recall@8 vs.\ SNR showing that reliable top-$k$ selection
    ($>90$\% recall) requires SNR~$\gtrsim \SI{15}{dB}$.
  }
  \label{fig:snr_analysis}
\end{figure}

\Cref{fig:snr_analysis} summarises the interplay between signature
dimension, photodetector SNR, and ranking accuracy across the
operating envelope of \prism{}.

\paragraph{Balanced photodetection noise.}
In the balanced configuration, each channel uses two photodetectors
measuring through-port and drop-port signals independently~\cite{Tait2017}.
Shot noise from both PDs adds in quadrature:
$\sigma_I^2 = 2e(I_{\text{through}} + I_{\text{drop}})\Delta f$.
Since $I_{\text{through}} + I_{\text{drop}} = \mathcal{R} P_0$ (power conservation),
the total shot noise is weight-independent, simplifying the noise analysis.
The factor of $\sqrt{2}$ increase in noise is offset by the doubled
signal dynamic range ($[-1,+1]$ vs $[0,1]$).

%% file: sections/system_evaluation.tex
\section{System-Level Evaluation}
\label{sec:system_eval}
\label{sec:software}

This section evaluates the complete \prism{} pipeline from
algorithmic profiling through end-to-end validation.
We first profile retrieval heads across two LLM families
(\cref{sec:sw_rh}), then evaluate block signature
design and recall (\cref{sec:sw_signature}), and validate
downstream accuracy via Needle-in-a-Haystack experiments
with MRR-simulated block selection (\cref{sec:sw_niah}).

\subsection{Retrieval-Head Analysis}
\label{sec:sw_rh}
\label{sec:sw_retrieval}
\label{sec:eval_retrieval}

\paragraph{Models and datasets.}
We profile two representative open-weight LLMs:
Qwen2.5-7B-Instruct~\cite{Qwen2_2024} ($L=28$, $H=28$,
$d_h=128$, GQA~\cite{Ainslie2023GQA} with $H_{\text{KV}}=4$; total 112 KV heads) and
Qwen3-8B~\cite{Qwen3_2025} ($L=36$, $H=32$,
$d_h=128$, GQA with $H_{\text{KV}}=8$; total 288 KV heads).
Qwen2.5-7B supports context lengths of at least \num{128000} tokens;
Qwen3-8B supports up to \num{32000} tokens.
We compute retrieval ratios $R_h^{(\ell,h)}$ (\cref{eq:retrieval_ratio})
on a calibration set of 2--3 random token sequences per context length,
with $w = 256$ as the local window size.
The retrieval ratio is measured using a two-step procedure: SDPA-based
prefill followed by eager last-token attention extraction.
All experiments are run on an NVIDIA RTX 5880 (48\,GB VRAM) for
bf16 models, and an NVIDIA RTX 5070 (12\,GB) for 4-bit quantized
variants.
We additionally verify consistency between bf16 and 4-bit quantized
Qwen2.5-7B, finding that quantization does not substantially alter
retrieval head identification (e.g., 91.1\% vs.\ 92.0\% at 8K context
for bf16 and 4-bit, respectively).

\begin{figure}[!htb]
  \centering
  \includegraphics[width=\columnwidth]{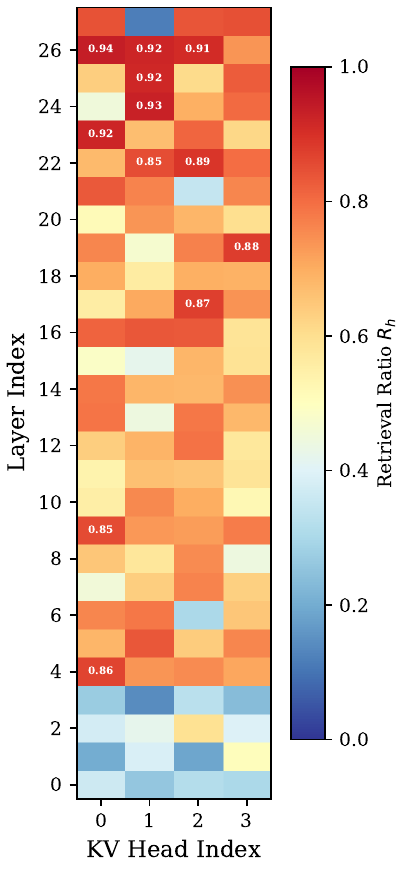}
  \caption{%
    Retrieval ratio $R_h^{(\ell,h)}$ for each KV head across
    all layers.
    Heads with $R_h > 0.3$ (dashed line) are classified as
    retrieval heads.
    (a)~Qwen2.5-7B: 102/112 heads are retrieval heads
    (91.1\%) at 8K context.
    (b)~Qwen3-8B: 258/288 heads are retrieval heads
    (89.6\%) at 8K context.
  }
  \label{fig:retrieval_heatmap}
\end{figure}

\paragraph{Results.}
\Cref{fig:retrieval_heatmap} shows the retrieval ratio heatmap
across all layers and heads for both models.

\Cref{tab:retrieval_vs_context} summarizes the retrieval head
fraction as a function of context length for both models.

\begin{table}[!htb]
  \centering
  \small
  \caption{%
    Retrieval head fraction at threshold $\tau = 0.3$ across context
    lengths. $R_h(\tau)$: percentage of KV heads with $R_h > \tau$.
    Mean $\bar{R}_h$: average retrieval ratio across all heads.
  }
  \label{tab:retrieval_vs_context}
  \begin{tabular}{@{}llccc@{}}
    \toprule
    Model & Context & $R_h(\tau{=}0.3)$ (\%) & Heads & Mean $\bar{R}_h$ \\
    \midrule
    \multirow{7}{*}{Qwen2.5-7B}
      & 2K   & 83.9 & 94/112  & 0.574 \\
      & 4K   & 83.0 & 93/112  & 0.560 \\
      & 8K   & 91.1 & 102/112 & 0.627 \\
      & 16K  & 92.9 & 104/112 & 0.639 \\
      & 32K  & 95.5 & 107/112 & 0.656 \\
      & 65K  & 92.0 & 103/112 & 0.633 \\
      & 128K & 98.2 & 110/112 & 0.796 \\
    \midrule
    \multirow{3}{*}{Qwen3-8B}
      & 2K  & 86.5 & 250/288 & 0.626 \\
      & 4K  & 88.2 & 254/288 & 0.652 \\
      & 8K  & 89.6 & 258/288 & 0.657 \\
    \bottomrule
  \end{tabular}
\end{table}

\begin{figure}[!htb]
  \centering
  \includegraphics[width=\columnwidth]{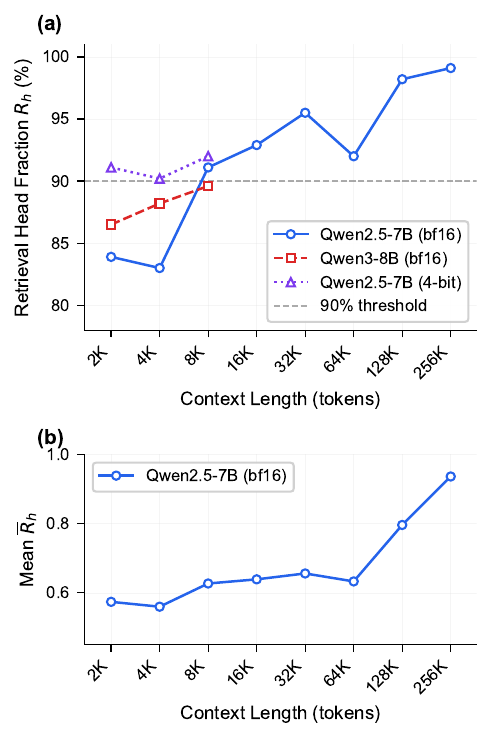}
  \caption{%
    Retrieval head statistics vs.\ context length.
    (a)~Retrieval head fraction $R_h(\tau{=}0.3)$ for
    Qwen2.5-7B (bf16 and 4-bit) and Qwen3-8B (bf16).
    The fraction exceeds 90\% for $n \geq 8$K and approaches
    99\% at 256K context.
    (b)~Mean retrieval ratio $\bar{R}_h$ for Qwen2.5-7B (bf16),
    showing that individual-head retrieval strength also
    increases with context length.
  }
  \label{fig:rh_context_scaling}
\end{figure}

\Cref{fig:rh_context_scaling} visualizes the retrieval head fraction
and mean retrieval ratio as a function of context length.

We observe the following patterns:
\begin{itemize}[leftmargin=*,itemsep=2pt]
  \item \textbf{Ubiquity of retrieval behavior.}
    At a threshold of $\tau = 0.3$, 91.1\% of KV heads in Qwen2.5-7B
    and 89.6\% in Qwen3-8B are retrieval heads at 8K context.
    This prevalence increases with context length: for Qwen2.5-7B,
    the fraction rises from 83.9\% at 2K to 98.2\% at 128K context,
    indicating that nearly all heads engage in long-range retrieval
    at long contexts.
    Note that at the more permissive $\tau = 0.1$ threshold used
    in~\cite{DuoAttention2024}, essentially 100\% of heads qualify
    as retrieval heads.
    The reported fraction is thus sensitive to the threshold choice:
    varying $\tau$ from 0.1 to 0.3 shifts the classified fraction
    from ${\sim}100\%$ to ${\sim}90\%$.
    The contrast with DuoAttention's 25--50\% retrieval fraction
    reflects both (i)~different models (Llama-2/Mistral vs.\ Qwen)
    and (ii)~DuoAttention's use of a learned gating function
    optimized on calibration data, which imposes a stricter
    criterion than a simple threshold on attention mass.
    In practice, the threshold can be tuned per deployment scenario
    to trade off between the number of heads served photonically
    and the complexity of the photonic accelerator.
  \item \textbf{Layer distribution.}
    The highest-scoring retrieval heads are concentrated in
    layers 14--26, with peak retrieval ratios exceeding 0.93.
  \item \textbf{GQA effect.}
    Because GQA shares KV heads across multiple query heads, the
    number of \emph{KV cache entries} requiring retrieval-style
    treatment is even smaller than the head count suggests.
\end{itemize}

The key implication for \prism{} is that the photonic accelerator
needs to serve the vast majority of KV heads---102 out of 112
for Qwen2.5-7B and 258 out of 288 for Qwen3-8B at 8K context.
However, GQA sharing means each KV head serves multiple query heads,
so the \emph{number of independent weight bank instances} required
equals the KV head count, not the query head count.

\subsection{Block Signature Design}
\label{sec:sw_signature}
\label{sec:eval_signature}
\label{sec:sw_recall}

We partition the KV cache into contiguous blocks of $B$ tokens
and compute a $d$-dimensional signature for each block~\cite{Indyk1998ANN}.
We evaluate mean-key and random projection signature methods
from \cref{sec:arch_signature} at block size $B = 128$ and
signature dimensions $d \in \{16, 32, 64, 128\}$,
using Qwen2.5-7B at context length $n = 4096$.
Our experiments identify $B = 128$ with $d = 32$ and mean-key
projection as the best configuration.
At the primary operating point $k{=}32$, \cref{tab:recall_params}
shows R@32 = 100\% at 8K context ($B{=}128$, 64 blocks), confirming
that the signature ranking correctly identifies all relevant blocks.
At 16K, R@32 drops to 57.5\%, yet downstream NIAH accuracy remains
100\% (\cref{tab:niah_mrr_extended}), indicating that task-critical
blocks are consistently ranked in the top-$k$ even when overall recall
is imperfect.
As a stress-test analysis at $k{=}8$, R@8 = 77.3\%
(R@2 = 31.3\%, R@4 = 50.0\%), confirming that useful ranking signal
persists even under aggressively small selection budgets.
Mean-key projection consistently outperforms random projection
across all tested dimensions, confirming that the natural key-space
geometry contains exploitable structure for block ranking.

\paragraph{Why mean-key and random projections?}
We focus on mean-key and random projection signatures because they are
model-agnostic and require no training, matching our goal of a
general-purpose photonic hardware interface.
Learned projections (e.g., trained linear maps optimized for recall)
could improve signature quality but would require per-model
fine-tuning and hardware-aware training, which we leave to future
work (\cref{sec:disc_future}).

\begin{figure}[!htb]
  \centering
  \includegraphics[width=\columnwidth]{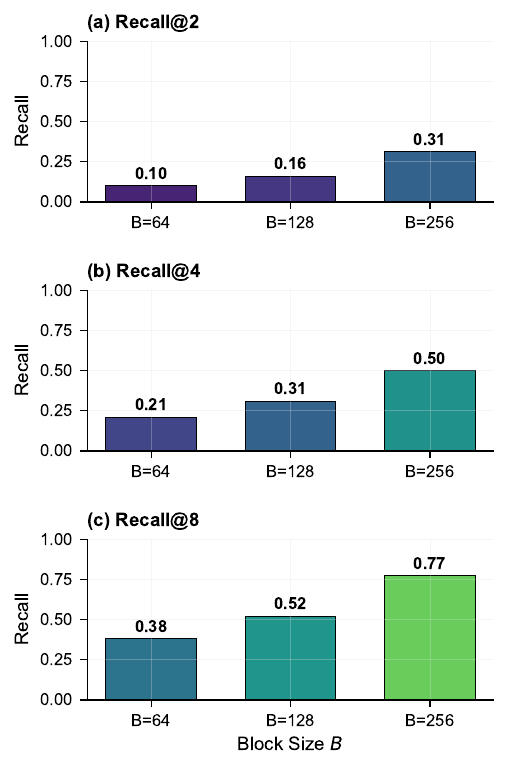}
  \caption{%
    Recall@$k$ as a function of signature dimension $d$
    for different signature methods.
    Block size $B = 128$, $k = 8$ (stress-test setting).
    Mean-key projection consistently outperforms random projection,
    achieving 77.3\% recall@8 at $d = 32$.
    At the primary operating point $k{=}32$, recall reaches 100\%
    at 8K context (\cref{tab:recall_params}).
  }
  \label{fig:recall_vs_dim}
\end{figure}

\paragraph{Signed weight encoding.}
The add-drop MRR configuration with balanced photodetection enables
direct encoding of signed weights $w \in [-1,\,+1]$, eliminating the
non-negative constraint of through-port-only architectures.
Compared to ReLU projection (which discards sign information, losing
${\sim}50\%$ of the signature variance for zero-mean Gaussian signatures),
balanced photodetection preserves the full signed inner product.
Our simulations show that signed encoding improves Recall@8 by
${\sim}87\%$ relative to ReLU projection at $d = 32$
(Supplementary \cref{fig:supp_signed_recall}).

\paragraph{Recall metric.}
We define recall@$k$~\cite{Manning2008IR} as the fraction of the true top-$k$ blocks
(by exact query--key inner product) that appear in the \prism{}-selected
top-$k$ blocks:
\begin{equation}
  \text{Recall@}k = \frac{|\mathcal{S}_{\text{PRISM}} \cap
    \mathcal{S}_{\text{exact}}|}{k},
  \label{eq:recall}
\end{equation}
where $\mathcal{S}_{\text{PRISM}}$ and $\mathcal{S}_{\text{exact}}$
are the sets of top-$k$ block indices selected by \prism{} and by
exact computation, respectively.

\begin{table}[!htb]
  \centering
  \small
  \caption{%
    Recall@$k$ for \prism{} block selection across context lengths.
    Qwen2.5-7B, $B{=}128$, $d{=}32$, mean-key projection.
    Values averaged over 15 (layer, head) pairs.
  }
  \label{tab:recall_params}
  \begin{tabular}{@{}cccccc@{}}
    \toprule
    $n$ & Blocks & R@8 (\%) & R@16 (\%) & R@32 (\%) & NIAH (\%) \\
    \midrule
    4K   & 16  & 46.7 & 100  & ---$^*$  & 100 \\
    8K   & 32  & 29.2 & 55.8 & 100  & 100 \\
    16K  & 64  & 26.7 & 41.7 & 57.5 & 100 \\
    32K  & 128 & \multicolumn{3}{c}{(OOM$^\dagger$)} & 100 \\
    64K  & 256 & \multicolumn{3}{c}{---}  & 100 \\
    \bottomrule
  \end{tabular}
  \\[3pt]
  {\footnotesize $^*$Only 16 blocks at 4K; $k{=}32$ exceeds total.\\
   $^\dagger$Eager attention OOM at 32K; NIAH uses SDPA (no attention matrix).}
\end{table}

\paragraph{Traffic reduction.}
At the primary operating point $k{=}32$, the traffic ratio is
$kB/n = 32 \times 128 / n$.
At 128K tokens ($N{=}1024$ blocks), PRISM selects $k{=}32$ of
$N{=}1024$ blocks, yielding a $N/k = 1024/32 = 32\times$ traffic
reduction (3.1\% traffic).
At 1M tokens ($N \approx 7812$ blocks), the reduction grows to
$N/k \approx 7812/32 \approx 244\times$ (0.41\% traffic), though
model accuracy at such lengths remains model-dependent.
Under the stress-test setting $k{=}8$, the reduction reaches
$128\times$ at 128K and projects to ${\sim}977\times$ at 1M tokens
(see Supplementary \cref{fig:supp_traffic_reduction}).

\subsection{NIAH Accuracy Under Hardware Impairments}
\label{sec:sw_niah}
\label{sec:niah_validation}
\label{sec:hw_niah}

To validate that the MRR-impaired block selection preserves
end-to-end language model performance, we integrate the MRR array
simulator into Qwen2.5-7B~\cite{Qwen2_2024} and evaluate on the
Needle-in-a-Haystack (NIAH) benchmark~\cite{Wu2025RetrievalHead,Kamradt2023NIAH}.

For each decode step, block signatures (mean-key, $d=32$) are
processed through the MRR simulator to select the top-$k$ blocks.
Retrieval heads ($R_h > 0.3$; \cref{tab:retrieval_vs_context}) use
MRR-selected blocks plus a 256-token recent window;
streaming heads retain full attention.
We test four MRR configurations: (i)~ideal (floating-point inner
product), (ii)~5-bit/\SI{20}{pm} drift (nominal),
(iii)~4-bit/\SI{30}{pm} drift (pessimistic), and
(iv)~5-bit/\SI{10}{pm} drift (optimistic).

\begin{table}[!htb]
  \centering
  \small
  \caption{%
    NIAH accuracy (\%) with MRR-integrated block selection
    (Qwen2.5-7B, 11 positions, $k = 8$ stress-test setting).
  }
  \label{tab:niah_mrr}
  \begin{tabular}{@{}lccc@{}}
    \toprule
    Configuration & 2K & 4K & 8K \\
    \midrule
    Full attention   & 90.9 & 100.0 & 100.0 \\
    Ideal select     & 90.9 & 100.0 & 100.0 \\
    5-bit, 20\,pm    & 90.9 & 100.0 & 100.0 \\
    4-bit, 30\,pm    & 90.9 & 100.0 & 100.0 \\
    5-bit, 10\,pm    & 90.9 & 100.0 & 100.0 \\
    \bottomrule
  \end{tabular}
\end{table}

\Cref{tab:niah_mrr} shows that all four MRR configurations---including
the worst-case 4-bit quantization with \SI{30}{pm} thermal
drift---achieve \emph{identical} NIAH accuracy to full attention
at all tested context lengths.
The single miss at 2K context (position 50\%) is a model-level
artifact unrelated to block selection.
These results demonstrate that the MRR impairments modelled in
\cref{sec:hw_impairments} do not degrade downstream task accuracy
for the block-selection ranking task.

To validate \prism{} across a wide range of context lengths, we
extend the evaluation using SDPA-based attention (Flash Attention)
with KV cache offloading to CPU RAM via \texttt{OffloadedCache}.
This enables experiments at context lengths from 4K to 128K on a
single GPU (RTX~5880, \SI{48}{\giga\byte}) with \SI{128}{\giga\byte}
system RAM.
We note that Qwen2.5-7B's native context window is 128K tokens;
at 128K, the base model's own accuracy degrades to 45.5\% on NIAH
(\cref{tab:niah_mrr_extended}), limiting meaningful evaluation
beyond 64K.
Extrapolation to longer contexts (e.g., 1M tokens via
YaRN~\cite{Yang2025Qwen1M} rope scaling) is technically feasible
for the photonic hardware, but model-level accuracy at such lengths
remains an open challenge independent of block selection.

For sparse evaluation, we employ \emph{physical token selection}:
rather than re-attending to all tokens with a mask, only the
tokens from the top-$k$ selected blocks and a recent window are
assembled into a compact input ($\sim$5K tokens), preserving
positional encoding via explicit \texttt{position\_ids}.
This approach mirrors the actual deployment scenario where only
selected KV blocks are fetched from memory.

\begin{figure}[!htb]
  \centering
  \includegraphics[width=\columnwidth]{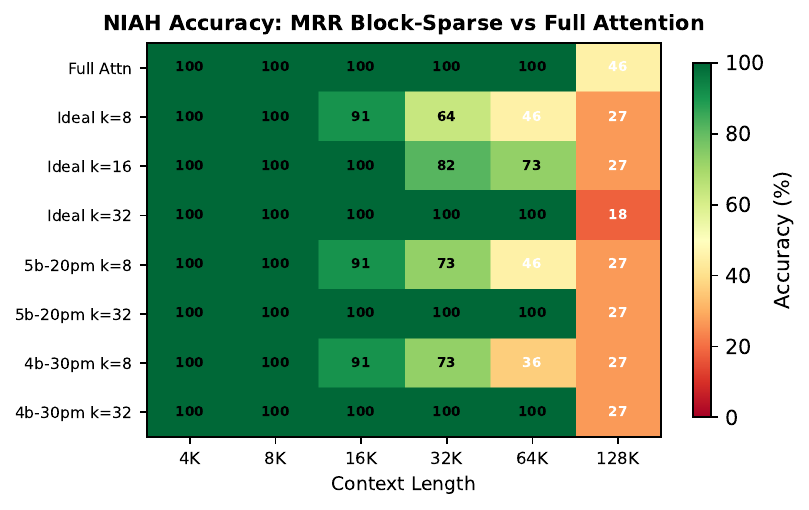}
  \caption{%
    NIAH accuracy heatmap across context lengths (4K--128K) and
    MRR configurations.
    At $k{=}32$, all MRR variants match full attention perfectly
    up to 64K.
    Under the $k{=}8$ stress test, accuracy degrades gracefully
    with context length but remains above 90\% at 16K.
    The 128K column shows model-intrinsic degradation
    (full attention itself drops to 45.5\%).
  }
  \label{fig:niah_heatmap}
\end{figure}

\Cref{tab:niah_mrr_extended} presents the extended NIAH results
across context lengths from 4K to 128K.
The full 2D NIAH heatmap (context length $\times$ needle depth,
10 positions per context) is shown in \cref{fig:niah_heatmap}.
At $k{=}32$ blocks ($B{=}128$), all MRR configurations achieve
\textbf{100\% accuracy from 4K through 64K}, perfectly matching
full attention.
At 128K, the base model itself degrades to 45.5\%---a known
limitation of Qwen2.5-7B's context window---making
sparse-vs-full comparison uninformative at this length.
Within the model's reliable operating range ($N \leq 64$K),
MRR block selection introduces \emph{zero} accuracy penalty
while reducing KV memory traffic by $16\times$ at 64K
($k \cdot B / n = 32 \times 128 / 65536 = 6.25$\%);
the reduction grows to $32\times$ (3.1\%) at 128K.

\begin{table}[!t]
  \centering
  \caption{%
    Extended NIAH accuracy (\%) with MRR block selection
    (Qwen2.5-7B, $B{=}128$, $d{=}32$, $k{=}32$,
    10 positions per context).
    At 128K the base model degrades ($\dagger$).
  }
  \label{tab:niah_mrr_extended}
  \begin{tabular}{@{}lcccccc@{}}
    \toprule
    Configuration & 4K & 8K & 16K & 32K & 64K & 128K$^\dagger$ \\
    \midrule
    Full attention   & 100 & 100 & 100 & 100 & 100 & 45.5 \\
    Ideal ($k{=}32$)     & 100 & 100 & 100 & 100 & 100 & 18.2$^\ddagger$ \\
    5-bit, 20\,pm    & 100 & 100 & 100 & 100 & 100 & 27.3$^\ddagger$ \\
    4-bit, 30\,pm    & 100 & 100 & 100 & 100 & 100 & 27.3$^\ddagger$ \\
    5-bit, 10\,pm    & 100 & 100 & 100 & 100 & 100 & 27.3$^\ddagger$ \\
    \bottomrule
  \end{tabular}
  \par\vspace{2pt}
  {\footnotesize $^\ddagger$At 128K,
    full attention itself degrades to 45.5\%; the apparent superiority of
    impaired configurations over ideal is within the ${\pm}9.1$\% sampling
    noise of 11 needle positions and is not statistically significant.}
\end{table}

%% file: sections/scaling_analysis.tex
\section{Photonic Scaling Analysis}
\label{sec:scaling}

We now analyze how the photonic engine scales to larger systems,
identifying constraints from WDM channel density, thermal power,
chip area, and time-multiplexed operation.

\subsection{MRR Integration Scaling}
\label{sec:scale_mrr}

The total MRR count in the \prism{} weight bank is
\begin{equation}
  N_{\text{MRR}} = d \times N,
  \label{eq:mrr_count}
\end{equation}
where $d$ is the number of WDM wavelength channels (signature
dimension) and $N$ is the number of parallel signature banks
(one per KV cache block).
For a context length of $n$ tokens with block size $B$,
$N = n / B$.

\Cref{tab:scaling_configs} lists representative configurations
spanning three orders of magnitude in MRR count.

\begin{table}[!htb]
  \centering
  \small
  \caption{%
    MRR count for representative \prism{} configurations.
    The rightmost column indicates the approximate context length
    supported at block size $B = 128$.
  }
  \label{tab:scaling_configs}
  \begin{tabular}{@{}rrrrl@{}}
    \toprule
    $d$ & $N$ & $N_{\text{MRR}}$ & Context & Feasibility \\
    \midrule
    32  & 256   & \num{8192}    & 32K   & Current TFLN \\
    64  & 1024  & \num{65536}   & 128K  & Near-term \\
    128 & 4096  & \num{524288}  & 512K  & Multi-chip \\
    \bottomrule
  \end{tabular}
\end{table}

Current photonic integration supports $10^4$--$10^5$ active devices
per die~\cite{Huang2020,Shekhar2024SiPhRoadmap}, placing the
$d{=}32$, $N{=}256$ configuration within demonstrated capability and
$d{=}64$, $N{=}1024$ at the near-term frontier.
The $d{=}128$, $N{=}4096$ configuration exceeds single-chip density,
requiring chiplet-based multi-chip modules~\cite{Seok2019WaferScale}
(\cref{sec:scale_area}).

\subsection{Thermal Power Budget and WDM Channel Limits}
\label{sec:scale_thermal}
\label{sec:scale_wdm}

On thermo-optic SOI platforms~\cite{Padmaraju2014}, each MRR requires
${\sim}\SI{2.5}{\milli\watt}$ of static heater power, yielding
aggregate budgets of \SI{20}{\watt} ($d{=}32$, $N{=}256$) to
\SI{164}{\watt} ($d{=}64$, $N{=}1024$)---approaching the
${\sim}\SI{200}{\watt}$ practical limit with active cooling.
On the TFLN platform, MRR tuning via the Pockels effect
($r_{33} = \SI{30.9}{pm/V}$) is capacitive with near-zero static
power ($<\SI{1}{\micro\watt}$ per ring from CMOS driver leakage):
\begin{equation}
  P_{\text{static}}^{\text{TFLN}} = d \times N \times P_{\text{leakage}}
  < d \times N \times \SI{1}{\micro\watt}.
  \label{eq:p_static_tfln}
\end{equation}
For the $d{=}64$, $N{=}1024$ configuration,
$P_{\text{static}} < \SI{0.07}{\watt}$---a ${\sim}2400\times$
reduction over SOI.
The switching energy (${\sim}\SI{5}{\femto\joule}$ per ring) yields
$<\SI{0.3}{\micro\watt}$ total switching power at typical decode
rates---negligible.
Residual thermal stabilization via TEC
(${\sim}\SI{1}{\watt}$ for a ${\sim}\SI{1}{\centi\meter^2}$ chip)
remains necessary but is orders of magnitude below SOI heater budgets.
TFLN's lower thermo-optic coefficient
($\text{dn/dT} \approx \SI{4e-5}{K^{-1}}$ vs.\
$\SI{1.8e-4}{K^{-1}}$ for Si) further reduces thermal crosstalk.

\paragraph{WDM channel limits.}
The signature dimension $d$ is constrained by the MRR free spectral
range (FSR)~\cite{Dong2016WDM}.
A single-FSR MRR ($R = \SI{20}{\micro\meter}$,
FSR $\approx \SI{8.3}{\nano\meter}$) supports only ${\sim}5$
channels at \SI{200}{\giga\hertz} spacing.
Vernier-coupled dual-ring filters extend the effective FSR to
${\sim}\SI{50}{\nano\meter}$ ($d \sim 30$); C+L band operation
(\SI{95}{\nano\meter}) enables $d \sim 60$.
Achieving $d = 128$ requires FSR extension with C+L+S band
operation (Supplementary Section~S3).

\subsection{Chip Area Estimation}
\label{sec:scale_area}

The $32 \times 256$ configuration (\num{8192} MRRs) fits on a single
${\sim}5 \times 5$\,mm$^2$ die; the $64 \times 1024$ configuration
requires multi-chip packaging or folded layouts.
Detailed area estimates are provided in Supplementary Section~S4.

\Cref{fig:scaling_analysis} summarizes the scaling trend.

\begin{figure*}[!tp]
  \centering
  \includegraphics[width=\textwidth]{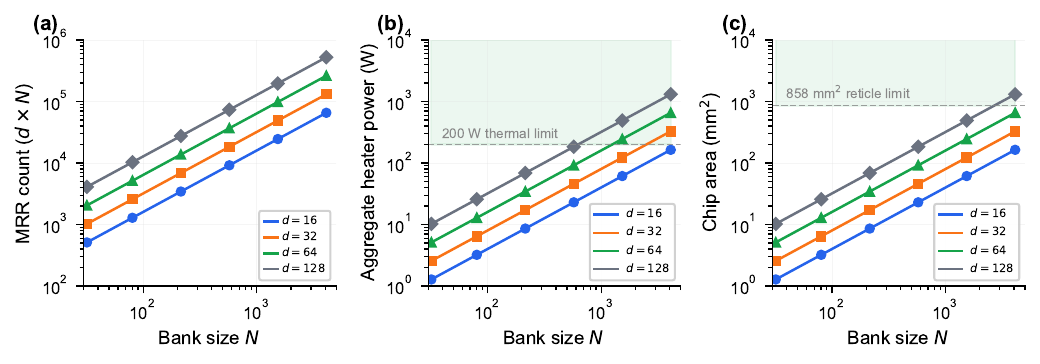}
  \caption{%
    \prism{} photonic scaling projections.
    MRR count, aggregate heater power (SOI), and estimated chip area
    as functions of the configuration parameters $d$ and $N$.
    The dashed horizontal lines indicate practical limits:
    \SI{200}{\watt} thermal dissipation (active cooling) and
    \SI{858}{\milli\meter\squared} single-reticle area.
    Configurations below both limits (shaded region) are realizable
    on a single photonic chip.
  }
  \label{fig:scaling_analysis}
\end{figure*}

\subsection{Time-Multiplexed Operation}
\label{sec:scale_timemux}

Area and power constraints can be relaxed by trading physical
parallelism for temporal reuse via time-multiplexed weight
programming~\cite{Bai2023PhotonicMux}.
The system deploys $N_{\text{phys}}$ physical rows and cycles
through $M$ weight configurations:
\begin{equation}
  N_{\text{logical}} = M \times N_{\text{phys}},
  \qquad M = \lceil N / N_{\text{phys}} \rceil.
  \label{eq:time_mux}
\end{equation}
On TFLN, EO reprogramming is sub-nanosecond
($t_{\text{reprogram}} \ll t_{\text{optical}}$), so the total
latency simplifies to
$t_{\text{total}} \approx M \times t_{\text{optical}}$.
Even at $M = 8$, the total latency
($\SI{80}{\nano\second}$) remains four orders of magnitude below
the GPU full-scan baseline (${\sim}\SI{200}{\micro\second}$)---a
fundamental advantage over thermo-optic SOI
($t_{\text{reprogram}} \sim \SI{10}{\micro\second}$).

\begin{figure*}[!tp]
  \centering
  \includegraphics[width=\textwidth]{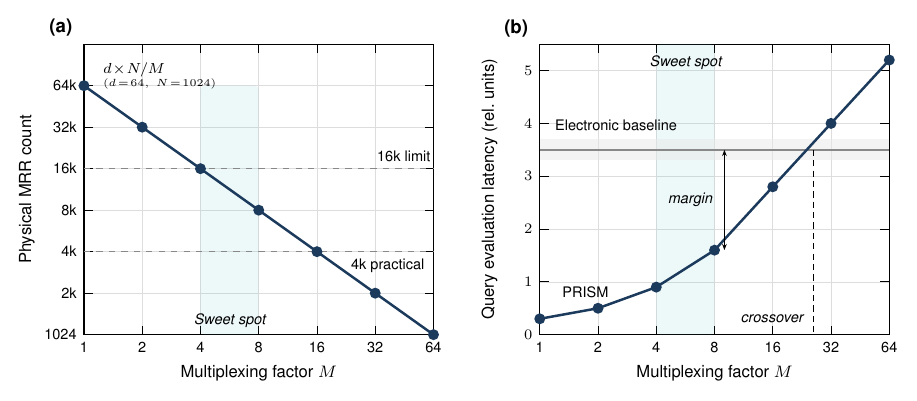}
  \caption{%
    Time-multiplexed \prism{} operation.
    (a)~Physical MRR count vs.\ multiplexing factor $M$
    ($d = 64$, $N = 1024$).
    (b)~Query evaluation latency vs.\ $M$, compared to electronic
    baselines.
    The shaded region indicates the sweet spot ($M = 4$--$8$).
  }
  \label{fig:time_multiplex}
\end{figure*}

For LLM decode at 128K+ context, $M = 4$--$8$ is a practical
sweet spot: it reduces physical MRR count by $4$--$8\times$ (to
\num{8192}--\num{16384}), keeps chip area within a single reticle,
and resolves the area scaling barrier of \cref{sec:scale_area},
making $d = 64$ realizable with current TFLN technology
(\cref{fig:time_multiplex}).

\subsection{Energy and Latency Crossover}
\label{sec:crossover}
\label{sec:cross_definition}
\label{sec:cross_results}
\label{sec:cross_scaling}
\label{sec:eval_crossover}

We define the crossover point $n^*$ as the context length at which
\prism{}-assisted decoding cost equals the electronic baseline.
The \prism{} cost (photonic selection energy ${\sim}\SI{931}{pJ}$
per query plus reduced GPU fetch) is compared against the GPU
full-scan cost (fetching all $N$ blocks via HBM).
On TFLN, near-zero static power means the selection cost is
dominated by dynamic components.
The full derivation is in Supplementary Section~S2.

\begin{figure}[!htb]
  \centering
  \includegraphics[width=\columnwidth]{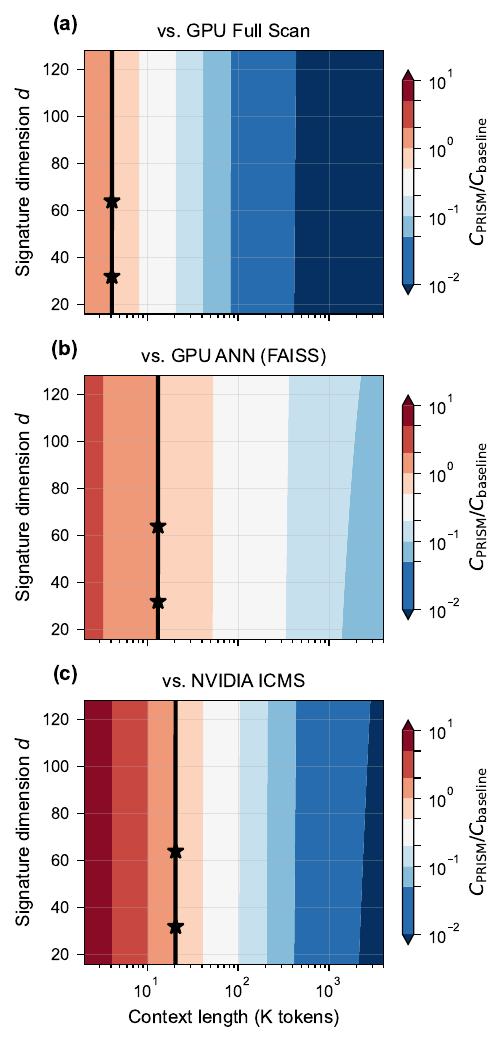}
  \caption{%
    Energy crossover map for \prism{} vs.\ electronic baselines.
    The crossover contour ($C_{\text{PRISM}} /
    C_{\text{baseline}} = 1$) shifts to shorter context lengths
    as $d$ decreases.
    (a)~vs.\ GPU full scan: practical benefit at $n^* \approx 4$K.
    (b)~vs.\ GPU ANN: $n^* \approx 2$K.
    (c)~vs.\ NVIDIA ICMS: $n^* \approx 4$K.
  }
  \label{fig:crossover_contour}
\end{figure}

\paragraph{Energy crossover.}
Against the GPU full scan (\cref{fig:crossover_contour}a),
the mathematical crossover occurs at $n^* < 1$K tokens
($d = 64$, $N_{\text{bank}} = 4$); practical benefit emerges at
$n \geq 4$K where traffic reduction exceeds $8\times$.
The per-query dynamic energy (${\sim}\SI{931}{pJ}$;
${\sim}\SI{9.9}{nJ}$ with amortized TEC) is five orders of
magnitude below the H100 fetch energy at 128K context
(${\sim}\SI{48}{\micro\joule}$).
This GPU baseline assumes a full-dimension scan ($d_h{=}128$).
A fairer comparison lets the GPU scan compressed $d{=}32$
signatures, reducing scan energy to
$E_{\text{scan}}^{\text{fair}} \approx \SI{12}{\micro\joule}$.
Even under this fairer comparison where the GPU scans compressed
$d{=}32$ signatures (${\sim}12\,\mu$J), \prism{} maintains a
four-order-of-magnitude advantage (${\sim}\SI{931}{pJ}$ vs.\
${\sim}\SI{12}{\micro\joule}$), preserving a comfortable crossover
margin.
On thermo-optic SOI, the ${\sim}\SI{164}{\watt}$ heater power
would place the crossover at $n^* \approx 4$K.
Against GPU ANN (FAISS IVF-PQ)~\cite{Johnson2021FAISS}
($O(\sqrt{N})$ scan reduction), the crossover is at
$n^* \approx 2$K.
Against NVIDIA ICMS (DPU with lower bandwidth than GPU HBM),
$n^* \approx 4$K based on estimated GTC 2024 specifications.

\paragraph{Latency crossover.}
The ${\sim}\SI{9}{\nano\second}$ photonic evaluation is orders of
magnitude below the ${\sim}\SI{5}{\micro\second}$ GPU scan,
so the selection step is effectively free in latency terms
($n^* \lesssim 4$K tokens).

\paragraph{Sensitivity.}
The dominant factors are:
(i)~dynamic power (${\sim}\SI{88}{\milli\watt}$, dominated by
TIAs/ADCs/DACs, negligible vs.\ electronic baselines);
(ii)~signature dimension $d$ (controls MRR count and area,
traded against recall quality, \cref{sec:sw_signature});
(iii)~bank count $N_{\text{bank}}$ (splitting loss vs.\
parallelism); and
(iv)~HBM bandwidth (HBM4 improvements shift the crossover to
longer contexts).

\paragraph{Scaling projections.}
The energy ratio $C_{\text{PRISM}} / C_{\text{GPU}}$ decreases as
${\sim}1/n$ because electronic scan cost grows linearly while
\prism{} selection cost is fixed.
At $n = \num{1000000}$ ($N \approx 8000$ blocks), the GPU reads
${\sim}\SI{1}{\mega\byte}$ of signatures per head per query;
\prism{} accommodates this with $N_{\text{bank}} = 8$ banks
($\num{512000}$ MRRs total).
In multi-agent scenarios, a single weight bank serves $A$ agents
simultaneously (only the query sketch changes), amortizing dynamic
power by $1/A$.

\begin{figure}[!htb]
  \centering
  \includegraphics[width=\columnwidth]{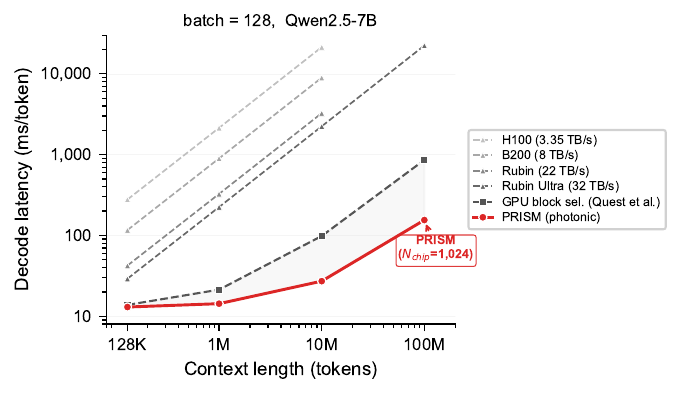}
  \caption{%
    Energy advantage vs.\ context length.
    The ratio decreases as ${\sim}1/n$; at \num{1000000} tokens
    ($B = 128$, $k{=}32$), the traffic reduction reaches
    $244\times$.
  }
  \label{fig:scaling_projection}
\end{figure}

%% file: sections/discussion.tex
\section{Discussion}
\label{sec:discussion}

\subsection{Limitations and Practical Considerations}
\label{sec:disc_limitations}
\label{sec:disc_scope}
\label{sec:disc_negative}
\label{sec:disc_multihead}
\label{sec:disc_fabrication}

All hardware results in this work are based on device-level
simulations with parameters extracted from FDTD and supplemented
by literature values; no physical prototype has been fabricated
or measured.
The impairment models, while grounded in FDTD simulation and
published device data, may not capture all fabrication-dependent
effects such as waveguide roughness variations, EO electrode
non-uniformity across a large array, and packaging-induced stress.
At $d = 64$ and $N = 1024$, the system requires \num{65536} MRRs;
systematic characterization of $>$\num{10000}-MRR arrays on TFLN
has not been reported, though recent progress in TFLN foundry
processes suggests that large-scale integration is
feasible~\cite{Hu2025TFLN_NatComm}.
Fabrication non-uniformity causes resonance wavelength variations
of $\sigma_{\lambda} \sim$\SIrange{0.5}{2}{\nano\meter} across a
wafer~\cite{Tait2016}, but on the TFLN platform EO tuning can
compensate via DC bias adjustment without static power penalty.
Residual thermal drift, while mitigated by lithium niobate's
${\sim}4\times$ lower thermo-optic coefficient compared to silicon,
still requires chip-level thermal stabilization
(${\sim}\SI{1}{\watt}$ TEC budget).
Including the TEC thermal stabilization overhead (amortized at
${>}\num{1000}$~queries/s throughput), the per-query energy rises from
the ${\sim}\SI{931}{pJ}$ dynamic-only figure to ${\sim}\SI{9.9}{nJ}$;
the $\SI{931}{pJ}$ value cited elsewhere in this paper refers to the
dynamic photonic pipeline alone.

\paragraph{Interface latency.}
The ${\sim}\SI{9}{ns}$ latency reported for \prism{} reflects the photonic pipeline alone
(DAC through top-$k$ selection) and does not include the host interface overhead.
A PCIe~5.0 round-trip (DMA setup and transfer) adds ${\sim}$\SIrange{1}{2}{\micro\second};
CXL-attached memory semantics reduce this to ${\sim}$\SIrange{200}{500}{\nano\second};
direct interposer or co-packaged integration would add only ${\sim}$\SIrange{10}{50}{\nano\second}.
Even with PCIe overhead, the total system latency of ${\sim}\SI{2}{\micro\second}$ remains
below the GPU full-scan latency (${\sim}\SI{5}{\micro\second}$), yielding a system-level
$2$--$3\times$ speedup.
Co-packaging---the long-term integration target---would preserve the
${\sim}100\times$ raw photonic advantage.
Speedup claims should therefore be interpreted as system-level $2$--$3\times$ with PCIe,
potentially $100\times$ with co-packaging.

\paragraph{Demonstrated vs.\ projected scale.}
To clarify the maturity of the MRR integration scales assumed in this work:
demonstrated TFLN arrays have reached ${\sim}$10--100 MRRs~\cite{Hu2025TFLN_NatComm},
while SOI platforms have demonstrated ${\sim}$\num{1000}--\num{10000}
MRRs~\cite{Huang2020}.
\prism{}'s ``current'' configuration (\num{8192}~MRRs at $d{=}32$, $N{=}256$)
is a \emph{projected} design point that extrapolates from these demonstrations;
the flagship configuration (\num{65536}~MRRs at $d{=}64$, $N{=}1024$) is also
projected and would likely require multi-chip or wafer-scale integration.

At $d = 64$ and $N = 1024$, the system requires \num{65536} individually
addressable voltage bias lines for fabrication-offset compensation of each MRR,
presenting a significant packaging and routing challenge that will require
advanced fan-out or interposer-based solutions.

The add-drop MRR configuration with balanced photodetection resolves
the sign limitation of through-port-only architectures.
The balanced differential photocurrent
$I_{\text{through}} - I_{\text{drop}}$ naturally encodes signed
weights in $[-1,+1]$, enabling true signed inner products without
ReLU projection or split encoding.
The trade-off is a doubling of the photodetector count (two PDs per
channel), but since PDs are orders of magnitude smaller than MRRs,
the area penalty is negligible.

The retrieval head classification threshold $\tau = 0.3$ used throughout
this work becomes less discriminative at longer contexts, where most heads
tend to exhibit high retrieval scores; DuoAttention's learned gating
identifies only 25--50\% of heads as retrieval heads.
The 90\%+ fraction reported here should therefore be interpreted as an
upper bound estimate at the evaluated context lengths.

For multi-head serving, GQA~\cite{Ainslie2023GQA} reduces the number of independent weight
bank instances from the retrieval head count (102 for Qwen2.5-7B)
to the KV head count ($H_{\text{KV}} = 4$), since block signatures
are derived from key vectors at KV-head granularity.
These 4 heads can be served by time-multiplexed reprogramming
(${\sim}\SI{4}{\nano\second}$ on TFLN, negligible vs.\ the
${\sim}\SI{5}{\micro\second}$ KV fetch) or by parallel replication
of 4 weight banks.
When the layer dimension is included, the full configuration space
is $H_{\text{KV}} \times L = 4 \times 28 = 112$ weight bank
instances per decode step.
Under time-multiplexing, this amounts to
$112 \times {\sim}\SI{1}{\nano\second} \approx \SI{112}{\nano\second}$
total reprogramming overhead---still ${\sim}45\times$ smaller than a
single KV block fetch (${\sim}\SI{5}{\micro\second}$) and therefore
negligible in the decode-step budget.
Alternatively, a layer-parallel deployment with 28 \prism{} chips
(one per layer, each serving 4 KV heads) would eliminate the layer
serialization entirely at the cost of additional chip area.

\subsection{Comparison with Related Approaches}
\label{sec:disc_infllm}
\label{sec:disc_icms}
\label{sec:sw_downstream}

The block-level top-$k$ selection mechanism at the core of \prism{}
builds on a strategy independently validated by several works:
Quest~\cite{Quest2024} preserves over 99\% of full-attention
accuracy on long-context benchmarks including NIAH up to 1M tokens,
DuoAttention~\cite{DuoAttention2024} maintains LongBench
performance within 1--2\% of full attention, and
InfLLM~\cite{InfLLM2024} and RocketKV~\cite{RocketKV2024} provide
additional evidence for block-level selection at long context.
\prism{}'s contribution is orthogonal: the key question is not
whether block selection preserves quality (answered affirmatively
above) but whether MRR-based analog computation introduces
sufficient error to degrade the selection.
Our NIAH results (\cref{sec:sw_niah}) confirm that it does not,
even under pessimistic hardware impairments.

Tian et al.'s photonic transformer chip (PTC)~\cite{Tian2025PTC}
demonstrates that coherent optical interference can implement full
transformer attention with high throughput ($>$200~POPS); however,
it targets dense attention computation rather than the coarse
block-selection task addressed by \prism{}, and its $O(n)$ memory
access scaling remains for long-context KV caches.

InfLLM is the most directly comparable system, as it offloads the
full KV cache to CPU RAM and retrieves blocks via electronic inner
products.
The key distinction is selection latency scaling: InfLLM's
selection time grows as $O(N)$ with the number of cached blocks,
while \prism{}'s photonic engine evaluates all $N$ similarities in
$O(1)$ optical transit time.
This advantage grows with context length---precisely the regime
where the KV cache bottleneck is most severe.

Relative to Quest~\cite{Quest2024} and RocketKV~\cite{RocketKV2024},
which perform block selection \emph{digitally} on the GPU,
\prism{} targets a different bottleneck: these methods reduce
\emph{compute} by pruning low-scoring KV blocks but still
require the GPU to read all block signatures from HBM
(costing $O(N)$ memory traffic per decode step).
\prism{} eliminates this signature scan entirely by offloading it to a
photonic co-processor with $O(1)$ latency and near-zero energy,
making it complementary---Quest- or RocketKV-style scoring
policies could be used to \emph{define} which blocks are
selected, while \prism{} accelerates the \emph{execution} of
that selection.
The GPU ANN baseline used in our crossover analysis
(FAISS IVF-PQ~\cite{Johnson2021FAISS}) represents a well-established
but not state-of-the-art GPU search library; more recent
GPU-accelerated ANN libraries (e.g., CAGRA, cuVS) may further reduce
the electronic baseline latency and energy, narrowing the crossover window.

NVIDIA's ICMS~\cite{NVIDIA_ICMS2026} addresses the complementary
\emph{capacity} problem (terabyte-scale flash-backed KV storage
with DPU-managed prefetch), while \prism{} solves the
\emph{selection} problem via photonic parallel inner products.
Note that the ICMS energy and bandwidth specifications used in our
comparisons are estimated from public announcements; no published
measurements are available, and actual performance may differ.
A natural integration would place \prism{} within or adjacent to
the ICMS, combining storage capacity with photonic selection speed.
The recently announced NVIDIA Rubin platform~\cite{Aubrey2026Rubin}
further underscores industry momentum toward dedicated KV cache
acceleration hardware, complementary to \prism{}'s photonic
approach.

\subsection{Outlook}
\label{sec:disc_future}

The immediate next step is fabrication of a small-scale TFLN MRR
prototype ($8 \times 8$ weight bank) to validate inner-product
accuracy under real device impairments and provide measured values
for parameters currently extracted from simulation.
Scaling to a full module ($d = 64$, $N = 256$) integrated with
GPU-based LLM inference would validate the crossover predictions
of \cref{sec:crossover}.
Integrating non-volatile weight storage (e.g., phase-change
trimming~\cite{Tossoun2024,Adya2025}) could further reduce
switching energy for quasi-static block
signatures~\cite{Fayza2025Photonics,Zhang2025PNN}.
More challenging benchmarks such as
SCBench~\cite{Li2025SCBench} and query-focused retrieval
analysis~\cite{Zhang2025QRHead} would strengthen confidence
in the robustness of photonically selected blocks beyond the
NIAH validation presented here.

\paragraph{Practical integration.}
A deployable \prism{} module would package the photonic chip, laser source,
and TEC onto a single substrate, offered in one of three form factors:
a PCIe add-in card for drop-in datacenter use, a CXL-attached device for
lower-latency memory-semantic access, or a co-packaged chiplet on an
interposer for maximum performance.
Integration with existing LLM serving stacks (e.g., vLLM, TensorRT-LLM)
would proceed via a block-index API: the host submits a query sketch and
receives ranked block indices, transparently replacing the software
signature-scan kernel.

\paragraph{Benchmark scope.}
NIAH is a retrieval-oriented benchmark that tests single-needle recall;
it does not exercise multi-hop reasoning, summarization, or other
long-context capabilities.
Our results therefore validate retrieval fidelity but not general
long-context quality.
We note, however, that the block selection mechanism is inherited from
Quest~\cite{Quest2024} and InfLLM~\cite{InfLLM2024}, which have been
validated on broader benchmarks (LongBench, $\infty$Bench);
\prism{}'s contribution is the photonic hardware mapping of this
selection, not the selection algorithm itself.

%% file: sections/conclusion.tex
\section{Conclusion}
\label{sec:conclusion}

We have presented \prism{}, a TFLN photonic similarity engine that
computes all $N$ block-selection inner products in $O(1)$ optical
latency via the broadcast-and-weight paradigm.
End-to-end NIAH evaluation confirms that MRR-selected block-sparse
attention preserves full-attention accuracy from 4K to 64K tokens
(within the model's native context window) under realistic hardware
impairments (4--5 bit weights, \SI{30}{pm} thermal drift), while
reducing KV cache traffic by $16\times$ at 64K context
($k{=}32$, $B{=}128$; $32\times$ at 128K).
At longer contexts (128K+), model-intrinsic accuracy degrades
independent of block selection; the photonic scaling analysis
nevertheless projects favorable energy and latency scaling to
million-token regimes as model context windows continue to expand.
The practical energy benefit emerges at $n \geq 4$K where block
selection yields meaningful traffic reduction, making \prism{}
favorable across virtually all practical context lengths.

Future work will proceed along three axes:
(i)~fabrication and characterization of an $8 \times 8$ TFLN MRR
weight bank to validate simulation predictions with measured device
parameters;
(ii)~scaling to a full $d{=}64$, $N{=}256$ module integrated with
GPU-based inference for end-to-end latency and energy measurements;
and (iii)~integration of non-volatile weight storage (e.g.,
phase-change trimming~\cite{Tossoun2024,Adya2025}) for write-once
signature programming, together with hardware-aware learned
projections and broader benchmarks such as
SCBench~\cite{Li2025SCBench}.
More broadly, photonic broadcast search may serve as a general
paradigm for similarity-search workloads in data centers---including
approximate nearest-neighbor retrieval, recommendation ranking, and
embedding lookup---wherever a single query must be compared against
a large, slowly changing set of stored vectors.

\section*{Disclosures}

The authors declare no conflicts of interest.

\section*{Data Availability}

Code and simulation data are available at
\url{https://github.com/hyoseokp/PRISM}~\cite{PRISM_code2025}.

%% file: sections/supplementary.tex
\appendix
\section*{Supplementary Information}
\setcounter{section}{0}
\renewcommand{\thesection}{S\arabic{section}}
\renewcommand{\theequation}{S\arabic{equation}}
\setcounter{equation}{0}
\renewcommand{\thefigure}{S\arabic{figure}}
\setcounter{figure}{0}
\renewcommand{\theHfigure}{supp.\arabic{figure}}

\section{Device Impairment Models}
\label{sec:supp_impairments}

This section provides the full mathematical models for the six
impairment sources that degrade the ideal inner-product
computation of \cref{eq:bw_inner_product}.

\paragraph{Weight quantization.}
MRR transmission is programmed via electro-optic tuning with finite
precision.
We model the quantized weight as
\begin{equation}
  \hat{w}_{n,j} = \frac{\text{round}(w_{n,j} \cdot 2^b)}{2^b},
  \label{eq:supp_quantization}
\end{equation}
where $b$ is the effective bit precision.
Values of $b = 4$--$8$ are considered, corresponding to
16--256 distinguishable transmission levels.

\paragraph{Thermal drift.}
After initial calibration, the MRR resonance wavelength drifts due
to ambient temperature fluctuations.
We model the drift as a Gaussian random walk:
\begin{equation}
  \Delta\lambda_0(t) = \sum_{i=1}^{t/\Delta t}
  \mathcal{N}(0, \sigma_{\text{drift}}^2),
  \label{eq:supp_thermal_drift}
\end{equation}
with $\sigma_{\text{drift}}$ chosen to produce a standard deviation
of \SIrange{0.01}{0.1}{\nano\meter} over a calibration interval
$T_{\text{cal}}$.
The resulting weight error is
\begin{equation}
  \Delta w = \left|\frac{\partial T}{\partial \lambda}\right|
  \Delta\lambda_0 \approx \frac{8Q^2 D_{\max}}{\lambda_0^2}
  \cdot \Delta\lambda_0,
  \label{eq:supp_weight_error_drift}
\end{equation}
evaluated at the operating point on the MRR Lorentzian.
Note that the i.i.d.\ Gaussian model above does not capture spatially
correlated drift (e.g., center-to-edge temperature gradients across the
chip), which could cause systematic bias in the inner-product scores
rather than zero-mean random noise; such gradients would require a
correlated noise model or per-region calibration.

\paragraph{Insertion loss.}
Each MRR introduces an off-resonance insertion loss
$\text{IL}_{\text{MRR}} \approx \SIrange{0.02}{0.05}{\deci\bel}$,
and the $1 \times N$ splitter contributes
$\text{IL}_{\text{split}}$ from \cref{eq:split_loss}.
The total channel loss is
\begin{equation}
  \text{IL}_{\text{total}} = \text{IL}_{\text{split}}
  + d \cdot \text{IL}_{\text{MRR}} + \text{IL}_{\text{wg}},
  \label{eq:supp_total_loss}
\end{equation}
where $\text{IL}_{\text{wg}}$ accounts for waveguide propagation
loss ($\sim$\SI{0.3}{\deci\bel/\centi\meter} for TFLN).
High insertion loss reduces the SNR at the photodetector and
increases the required laser power.

\paragraph{Photodetector noise.}
The photocurrent at each detector includes shot noise and thermal
noise:
\begin{equation}
  \sigma_I^2 = 2eI_{\text{ph}}\,\Delta f
  + \frac{4k_BT}{R_L}\,\Delta f + \text{NEP}^2 \cdot \Delta f,
  \label{eq:supp_detector_noise}
\end{equation}
where $I_{\text{ph}}$ is the signal photocurrent, $\Delta f$ is
the detection bandwidth, $R_L$ is the load resistance, and NEP
is the noise-equivalent power of the photodetector
($\sim \SI{10}{\pico\watt/\sqrt{Hz}}$ for Ge-on-Si)~\cite{Saleh2019Photonics,Personick1973}.
The noise introduces a random perturbation to the inner-product
score, potentially reordering the top-$k$ ranking.

\paragraph{MRR crosstalk.}
Adjacent MRRs on the same bus waveguide can exhibit spectral
overlap if the channel spacing is insufficient relative to the
MRR linewidth.
We model inter-channel crosstalk as an additive interference
with isolation of \SIrange{-15}{-30}{\deci\bel}:
\begin{equation}
  y_n = \sum_{j=1}^{d} w_{n,j}\,s_j + \sum_{j=1}^{d}
  \sum_{m \neq j} \chi_{j,m}\,w_{n,m}\,s_m,
  \label{eq:supp_crosstalk}
\end{equation}
where $\chi_{j,m}$ is the crosstalk coefficient from channel $m$
to channel $j$~\cite{Jayatilleka2016}.

\paragraph{Input DAC noise.}
The finite resolution and integral nonlinearity (INL) of the input
DACs contribute an additional noise floor on the query sketch values.
At $b_{\text{DAC}} = 6$ bits, the quantization noise standard
deviation is $\sigma_{\text{DAC}} = 2^{-b_{\text{DAC}}} / \sqrt{12}
\approx 0.0045$.


\begin{figure}[!htb]
  \centering
  \includegraphics[width=\columnwidth]{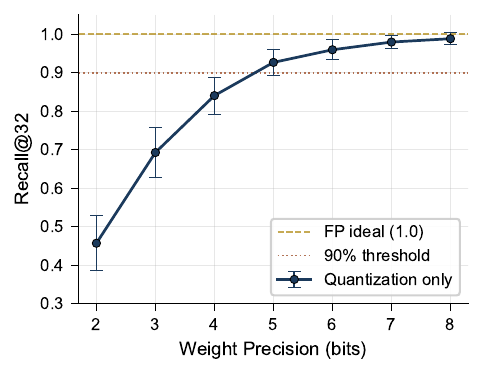}
  \caption{%
    Impact of weight quantization on recall.
    At $b = 6$ bits, recall degrades by less than 5\%
    from the floating-point ideal (Recall@8 = 0.960 at 6-bit).
    Adding thermal drift ($\sigma_{\text{th}} = 0.01$) and detector
    noise ($\sigma_{\text{det}} = 0.01$) degrades recall by an additional 5\%.
  }
  \label{fig:supp_recall_vs_precision}
\end{figure}

\begin{figure}[!htb]
  \centering
  \includegraphics[width=\columnwidth]{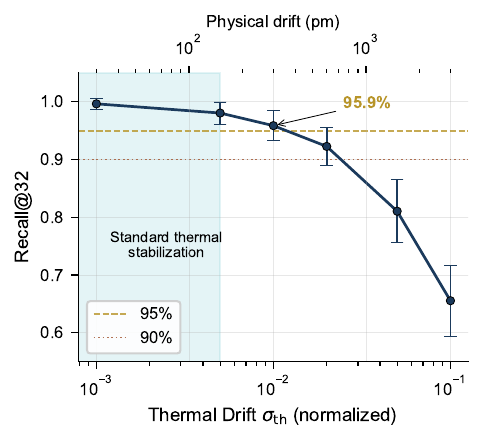}
  \caption{%
    Recall degradation as a function of thermal drift $\sigma$.
    Recall remains above 95\% for $\sigma \leq 0.005$
    (corresponding to $\sim$\SI{150}{pm} drift),
    achievable with standard thermal stabilization.
    At $\sigma = 0.01$ ($\sim$\SI{300}{pm}), recall is still 94.8\%.
  }
  \label{fig:supp_recall_vs_drift}
\end{figure}

\begin{figure*}[!tp]
  \centering
  \includegraphics[width=\textwidth]{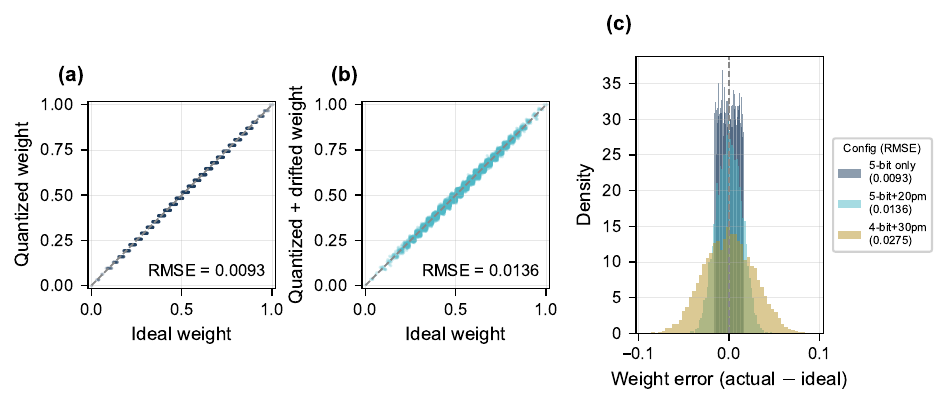}
  \caption{%
    Weight encoding fidelity under MRR impairments ($d = 32$, $N = 128$).
    (a)~Ideal vs.\ 5-bit quantised weights in $[0, 1]$: the staircase
    pattern shows 32 discrete levels with RMSE~$= 0.009$.
    (b)~Adding \SI{20}{pm} thermal drift and fabrication variation
    broadens the scatter (RMSE~$= 0.014$).
    (c)~Error histograms for three configurations: 5-bit only,
    5-bit with \SI{20}{pm} drift, and 4-bit with \SI{30}{pm} drift.
    Even the pessimistic case concentrates errors within $\pm 5$\%
    of the full weight range.
  }
  \label{fig:supp_weight_fidelity}
\end{figure*}

\begin{figure}[!htb]
  \centering
  \includegraphics[width=\columnwidth]{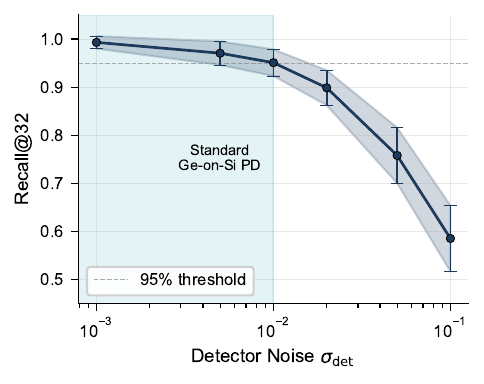}
  \caption{%
    Recall@8 degradation as a function of photodetector noise
    $\sigma_{\text{det}}$ ($d = 32$, $N = 500$, 100 trials).
    Recall remains above the 95\% threshold for
    $\sigma_{\text{det}} \leq 0.01$, achievable with standard
    Ge-on-Si photodetectors (NEP $\sim$ \SI{1}{pW/\sqrt{Hz}}).
  }
  \label{fig:supp_recall_vs_noise}
\end{figure}

\section{Crossover Derivation}
\label{sec:supp_crossover}

This section provides the full algebraic derivation of the
energy crossover point $n^*$ summarized in \cref{sec:crossover}.

The total decode cost per token for a retrieval head consists of
two terms:
\begin{equation}
  C_{\text{total}} = C_{\text{select}}(n) + C_{\text{fetch}}(n, k),
  \label{eq:supp_cost_total}
\end{equation}
where $C_{\text{select}}$ is the cost of determining which blocks
to fetch and $C_{\text{fetch}}$ is the cost of reading and computing
attention over the selected blocks.

For the \textbf{GPU full scan} baseline, no selection is needed
($k = N = n/B$), so
\begin{equation}
  C_{\text{GPU}} = C_{\text{fetch}}^{\text{GPU}}(n, n/B)
  = \frac{2\,d_h\,n\,b_{\text{prec}}}{\text{BW}_{\text{HBM}}},
  \label{eq:supp_cost_gpu}
\end{equation}
where $\text{BW}_{\text{HBM}}$ is the HBM bandwidth
($\sim$\SI{3.35}{TB/s} for H100 HBM3).

For \textbf{\prism{}}, the cost is
\begin{equation}
  C_{\text{PRISM}} = C_{\text{select}}^{\text{PRISM}}(n)
  + C_{\text{fetch}}^{\text{GPU}}(n, k\cdot B),
  \label{eq:supp_cost_prism}
\end{equation}
where $C_{\text{select}}^{\text{PRISM}}$ is the dynamic energy of
the photonic evaluation (laser, DAC, modulator, PD, ADC):
\begin{equation}
  C_{\text{select}}^{\text{PRISM}} \approx E_{\text{dynamic}},
  \label{eq:supp_cost_prism_select}
\end{equation}
Note that on the TFLN platform, the static MRR tuning power is
near zero (capacitive EO), so the selection cost is dominated
entirely by the dynamic components (${\sim}\SI{931}{pJ}$ per query,
\cref{tab:prism_energy}).
The fetch cost is reduced by the selection ratio $k/N$:
\begin{equation}
  C_{\text{fetch}}^{\text{GPU}}(n, k\cdot B)
  = \frac{2\,d_h\,k\,B\,b_{\text{prec}}}{\text{BW}_{\text{HBM}}}.
  \label{eq:supp_cost_prism_fetch}
\end{equation}

The crossover occurs when $C_{\text{PRISM}} < C_{\text{GPU}}$,
i.e., when the memory bandwidth saved by not fetching $(N - k)$
blocks exceeds the cost of operating the photonic selector.

Setting $C_{\text{PRISM}} = C_{\text{GPU}}$ and solving for $n$
yields:
\begin{equation}
  n^* = \frac{C_{\text{select}}^{\text{PRISM}} \cdot
  \text{BW}_{\text{HBM}}}{2\,d_h\,b_{\text{prec}}\,(1 - k\,B/n^*)},
  \label{eq:supp_crossover_implicit}
\end{equation}
which must be solved self-consistently since $k$ and $N = n/B$
both depend on $n$.
In practice, $k$ is a fixed parameter (e.g., $k = 32$), so the
selection ratio $k\,B/n \to 0$ as $n \to \infty$, and the crossover
simplifies to
\begin{equation}
  n^* \approx \frac{C_{\text{select}}^{\text{PRISM}} \cdot
  \text{BW}_{\text{HBM}}}{2\,d_h\,b_{\text{prec}}}.
  \label{eq:supp_crossover_approx}
\end{equation}

\section{WDM Channel Limits}
\label{sec:supp_wdm}

The maximum number of WDM channels $d$ is constrained by the MRR
free spectral range (FSR), the available optical bandwidth, and
inter-channel crosstalk.

\paragraph{Single-FSR constraint.}
For a TFLN MRR with radius $R = \SI{20}{\micro\meter}$, the FSR is
approximately \SI{8.3}{\nano\meter} (at $\lambda_0 = \SI{1550}{\nano\meter}$,
as in \cref{tab:device_params}).
At a channel spacing of $\Delta\lambda_{\text{ch}} = \SI{1.6}{\nano\meter}$
(\SI{200}{\giga\hertz} on the ITU grid)~\cite{ITU2020G694}, the maximum number of
non-aliased channels within one FSR is
\begin{equation}
  d_{\max}^{\text{(1-FSR)}} = \left\lfloor \frac{\text{FSR}}
  {\Delta\lambda_{\text{ch}}} \right\rfloor
  = \left\lfloor \frac{8.3}{1.6} \right\rfloor = 5.
  \label{eq:supp_d_max_fsr}
\end{equation}
This is clearly insufficient for the $d = 32$--$128$ range targeted
by \prism{}.

\paragraph{FSR extension techniques.}
Vernier-coupled dual-ring filters or cascaded MRRs with slightly
different radii can extend the effective FSR to
$\sim$\SI{50}{\nano\meter} or more~\cite{Tait2017,Boeck2010Vernier}, limited by the
least common multiple of the individual FSRs.
With an extended FSR of \SI{50}{\nano\meter} and
$\Delta\lambda_{\text{ch}} = \SI{1.6}{\nano\meter}$:
\begin{equation}
  d_{\max}^{\text{(Vernier)}} = \left\lfloor \frac{50}{1.6}
  \right\rfloor \approx 30.
  \label{eq:supp_d_max_vernier}
\end{equation}

\paragraph{Band-limited operation.}
The usable optical bandwidth depends on the operating band:
\begin{itemize}[leftmargin=*,itemsep=2pt]
  \item \textbf{C-band} (1530--\SI{1565}{\nano\meter}):
    \SI{35}{\nano\meter} $\rightarrow$ practical limit
    $d \sim 20$--$30$ without FSR extension;
  \item \textbf{C+L band} (1530--\SI{1625}{\nano\meter}):
    \SI{95}{\nano\meter} $\rightarrow$ $d \sim 60$.
\end{itemize}
For $d > 60$, extending to the S-band or using \SI{0.8}{\nano\meter}
(\SI{100}{\giga\hertz}) channel spacing is necessary, at the cost
of tighter crosstalk margins.
In practice, achieving $d = 128$ requires both FSR extension and
C+L+S band operation, representing a more aggressive photonic
design point.

\subsection{Balanced Photodetection for Signed Weights}
\label{sec:supp_balanced_pd}

The add-drop MRR configuration provides both through-port transmission
$T_{\text{through}}(\Delta\lambda)$ and complementary drop-port transmission
$T_{\text{drop}}(\Delta\lambda) = 1 - T_{\text{through}}(\Delta\lambda)$
(for a lossless ring).
A balanced photodetector pair measures the differential signal:
\begin{equation}
  w = T_{\text{through}} - T_{\text{drop}} = 2\,T_{\text{through}} - 1 \in [-1,\,+1].
\end{equation}
This mapping naturally encodes signed weights without doubling the MRR
count (as required by split encoding) or discarding sign information
(as in ReLU projection).
The noise model for balanced detection yields shot noise variance
$\sigma^2 = 2e\mathcal{R}P_0\Delta f$, independent of the programmed
weight, since power is conserved: $P_{\text{through}} + P_{\text{drop}} = P_0$~\cite{Saleh2019Photonics}.

\section{Multi-Head Serving with GQA}
\label{sec:supp_multihead}

A natural concern is whether serving all retrieval heads requires
replicating the weight bank for each head.
In Qwen2.5-7B, 102 out of 112 KV heads are retrieval heads at 8K
context (\cref{tab:retrieval_vs_context}).
However, GQA~\cite{Ainslie2023GQA} reduces the number of \emph{independent} KV heads to
$H_{\text{KV}} = 4$---each KV head is shared across $H/H_{\text{KV}} = 7$
query heads.
The weight bank stores block \emph{signatures} derived from key vectors,
so it operates at the KV-head granularity, not the query-head granularity.
This means only 4 independent weight bank configurations are needed
per layer, not 102.

Two strategies can serve these 4 KV heads:
\begin{enumerate}[leftmargin=*,itemsep=2pt]
  \item \textbf{Time-multiplexed reuse.}
    A single weight bank is reprogrammed sequentially for each of the
    4 KV heads.
    On the TFLN platform, EO tuning settles in ${\sim}\SI{1}{\nano\second}$
    (RC-limited), so reprogramming 4 heads adds only
    ${\sim}\SI{4}{\nano\second}$---negligible compared to the
    subsequent KV block fetch (${\sim}\SI{5}{\micro\second}$).
  \item \textbf{Parallel replication.}
    Four weight banks are deployed in parallel, one per KV head.
    This requires $4 \times d \times N = 4 \times 64 \times 1024
    = \num{262144}$ MRRs total---a $4\times$ increase over the
    single-head case but within the scalability limits discussed
    in \cref{sec:disc_fabrication}.
\end{enumerate}
GQA thus reduces the multi-head serving problem from 102 independent
banks (one per retrieval head) to just 4, making both time-multiplexed
and spatially-parallel approaches practical.
For Qwen3-8B ($H_{\text{KV}} = 8$), the same argument applies with
8 KV heads, still far below the 258 retrieval heads.

\section{Additional Figures}

\begin{figure}[!htb]
  \centering
  \includegraphics[width=\columnwidth]{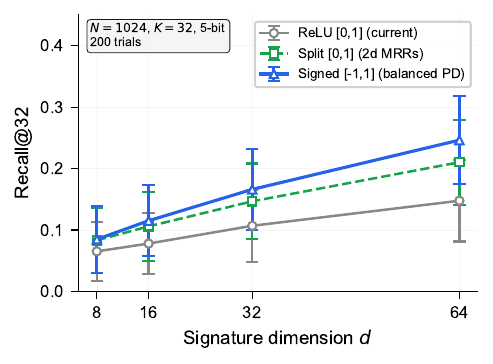}
  \caption{%
    Signed vs.\ unsigned recall comparison.
    Balanced photodetection (signed $[-1,+1]$) consistently outperforms
    ReLU projection (unsigned $[0,1]$) and split encoding across all
    signature dimensions $d$.
  }
  \label{fig:supp_signed_recall}
\end{figure}

\begin{figure}[!htb]
  \centering
  \includegraphics[width=\columnwidth]{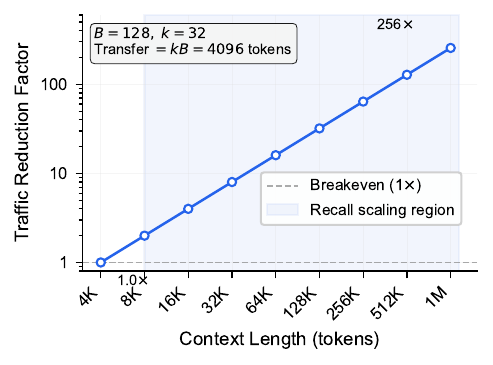}
  \caption{%
    KV cache traffic reduction factor $N/k$ as a function of context
    length for different $k$ values.
  }
  \label{fig:supp_traffic_reduction}
\end{figure}